\documentclass[11pt,letterpaper,preprint,amsmath,amssymb,tightenlines]{article}
\usepackage{jheppub}

\usepackage[english]{babel}
\usepackage{graphicx}
\usepackage{bm}
\usepackage{caption}
\usepackage{subfig}
\usepackage[countmax]{subfloat}
\usepackage{xspace}
\usepackage{hyperref}
\usepackage{verbatim}
\usepackage[normalem]{ulem}

\usepackage{multirow}

\usepackage{amssymb}
\usepackage{amsmath}
\usepackage{cancel}
\usepackage{xcolor}
\usepackage{slashed}
\usepackage{epstopdf} 
\usepackage{mathtools}

\usepackage{pdfsync}
\graphicspath{{figures/}} 
\def\nn{{\nonumber}}

\def\cO{\mathcal{O}}
\def\cP{\mathcal{P}}

\allowdisplaybreaks
\newcommand{\zero}{{(0)}}

\newcommand{\Tr}{\mathrm{Tr}}
\newcommand{\tr}{\operatorname{tr}}

\definecolor{darkyellow}{rgb}{0.5, 0.5, 0.0}
\definecolor{darkpurple}{rgb}{0.5, 0.2, 0.8}
\definecolor{darkblue}{rgb}{0.0, 0.0, 0.8}
\definecolor{darkgreen}{rgb}{0.0, 0.4, 0.0}
\definecolor{darkred}{rgb}{0.5, 0.0, 0.0}

\hypersetup{
    linktocpage,
     colorlinks,
     citecolor=darkgreen,
     linkcolor= darkgreen,
     urlcolor=darkgreen
}

\newcommand{\eq}[1]{Eq.~\eqref{eq:#1}}
\newcommand{\eqs}[2]{Eqs.~\eqref{eq:#1} and \eqref{eq:#2}} 

\renewcommand{\sec}[1]{Sec.~\ref{sec:#1}}

\newcommand{\fig}[1]{Fig.~\ref{fig:#1}}

\newcommand{\app}[1]{App.~\ref{app:#1}}

\DeclareRobustCommand{\Refs}[1]{Refs.~\cite{#1}}

\newcommand{\df}{\mathrm{d}}

\newcommand{\bt}{\vec{b}_T}
\newcommand\bn{{\bar n}}

\newcommand{\Mae}[3]{\bigl\langle#1\bigr\rvert#2\bigr\rvert#3\bigr\rangle}

\def\cB{\mathcal{B}}

\def\cJ{\mathcal{J}}

\def\cO{\mathcal{O}}

\def\cP{\mathcal{P}}
\def\cJ{\mathcal{J}}

\def\tr{{\rm tr}}

\def\bT{\mathbf{T}}

\newcommand{\brk}{\nonumber\\&}
\newcommand{\nbrk}{\nonumber\\}

\newcommand{\gsoft}{\gamma^{s}}

\def\ugr{\gamma^{r}}

\title{The Three-Point Energy Correlator in the Coplanar Limit}

\author[1]{Anjie Gao,}
\author[2]{Tong-Zhi Yang,}
\author[3]{and Xiaoyuan Zhang}

\affiliation[1]{Center for Theoretical Physics, Massachusetts Institute of Technology, Cambridge, MA 02139, USA}
\affiliation[2]{Physik-Institut, Universit\"at Z\"urich, Winterthurerstrasse 190, CH-8057 Z\"urich, Switzerland}
\affiliation[3]{Department of Physics, Harvard University, Cambridge, MA 02138, USA}

\emailAdd{anjiegao@mit.edu}
\emailAdd{toyang@physik.uzh.ch}
\emailAdd{xiaoyuanzhang@g.harvard.edu}

\preprint{\vbox{%
\hbox{MIT-CTP 5807}
\hbox{ZU-TH 54/24}
}
}
\abstract{
Energy correlators are a type of observables that measure how energy is distributed across multiple detectors as a function of the angles between pairs of detectors. In this paper, we study the three-point energy correlator (EEEC) at lepton colliders in the three-particle near-to-plane (coplanar) limit. The leading-power contribution in this limit is governed by the three-jet (trijet) configuration.
We introduce a new approach by projecting the EEEC onto the volume of the parallelepiped formed by the unit vectors aligned with three detected final-state particles. 
Analogous to the back-to-back limit of the two-point energy correlator probing the dijet configuration, the small-volume limit of the EEEC probes the trijet configuration.
We derive a transverse momentum dependent (TMD) based factorization theorem that captures the soft and collinear logarithms in the coplanar limit,
which enables us to achieve the next-to-next-to-next-to-leading logarithm (N$^3$LL) resummation. To our knowledge, this is the first N$^3$LL result for a trijet event shape.
Additionally, we demonstrate that a similar factorization theorem can be applied to the fully differential EEEC in the three-particle coplanar limit, which provides a clean environment for studying different coplanar trijet shapes.
}

\begin{document}
\maketitle
\flushbottom
\newpage

\section{Introduction}

The energy-energy correlation (EEC), which measures the energy deposited in two fixed detectors has recently attracted lots of attention in both perturbative QCD and collider physics.  As pointed out in Refs.~\cite{Basham:1978zq,Basham:1978bw}, the EEC is a nice QCD observable because the soft gluon divergence is suppressed by the energy weights and the collinear divergence is regulated by the angle between the two detectors. More recently, EEC has been generalized to a new family of QCD observables called
$n$-point {\it energy correlators}~\cite{Chen:2019bpb}, defined as weighted cross sections in terms of the angles between any two of the final-state particles,
\begin{equation}\label{eq:EC_def}
    \frac{\df\sigma}{\df x_{12}\cdots dx_{(n-1)n}} \equiv \sum_{m}\sum_{1\leq i_1,\cdots i_n \leq m}\int\! \df \sigma_{m}\times  \prod_{1\leq k \leq n}\frac{E_{i_k}}{Q} \prod_{1\leq j < l \leq n}\delta\left(x_{jl}-\frac{1-\cos\theta_{i_{j}i_{l}}}{2}\right)
    \,.
\end{equation}
Here $m$ is the number of final-state particles and $\df\sigma_m$ is the differential cross section for $m$ final-state particles given a specific theory. The sum over $i_1,\cdots i_n$ runs over all $n$ subsets of the final states and the sum over $m$ ensures the infrared safety due to Kinoshita-Lee-Nauenberg (KLN) theorem~\cite{Kinoshita:1962ur, Lee:1964is}. We use the $x_{ij}$ variables to label the angles between the detector $i$ and $j$, which goes to 0 (or 1) when two detectors are collinear (or back-to-back). For $n=2$, this definition reduces to the traditional EEC. For $n=3$, namely the three-point energy correlator (EEEC), the distribution encodes a non-trivial shape dependence that is associated with the QCD dynamics. This allows us to not only look at the shape of jets produced at colliders, but also probe the substructure inside a single jet.

For the past few decades, people have been dedicated to calculating scattering amplitudes at higher loops and understanding their mathematical structure. However, there are very few fixed-order data for the cross-section or collider observables. Monte Carlo simulations are preferred because the cuts from either phase space or jet algorithms make the integrals very complicated. Among these observables, energy correlators turn out to be the {\it simplest} one for analytic computation. For example, EEC has been computed to next-to-leading order (NLO) in QCD~\cite{Dixon:2018qgp,Luo:2019nig,Gao:2020vyx} and NNLO in $\mathcal{N}=4$ super Yang-Mills (SYM) theory~\cite{Belitsky:2013xxa,Belitsky:2013bja,Belitsky:2013ofa,Henn:2019gkr}. The leading order EEEC with full angle dependence is then calculated in both QCD~\cite{Yang:2022tgm,Yang:2024gcn} and $\mathcal{N}=4$ SYM~\cite{Yan:2022cye}. Very recently, the collinear limit of four-point energy correlator is also available~\cite{Chicherin:2024ifn}.
On one side, the available fixed-order data allows us to push the precision for Standard Model measurements and new physics searches.
On the other side, the analytic result reveals that there are various singular limits in the EEEC distribution, including triple collinear limit ($x_{12}, x_{13}, x_{23}\sim 0$ homogeneously), squeezed limit ($x_{12}\sim 0$ and $x_{13}, x_{23}\sim x$), coplanar limit (three final-state particles fall onto the same plane), etc. These limits are summarized and the fixed-order expansions are provided in Ref.~\cite{Yang:2022tgm}.

In perturbative quantum field theories, infrared divergences are always associated with large logarithms in some kinematic limits. Such logarithms could spoil the convergence of perturbation theory and one needs to resum them to all orders in $\alpha_s$ to retain the predictive power of perturbative expansions.
At $e^+e^-$ colliders, the EEC has been resummed to the next-to-next-to-leading logarithm (NNLL) accuracy~\cite{Dixon:2019uzg,Korchemsky:2019nzm} in the collinear limit which is known as the Dokshitzer-Gribov-Lipatov-Altarelli-Parisi (DGLAP) region, and to N${}^4$LL~\cite{COLLINS1981381,ELLIS198499,deFlorian:2004mp,Tulipant:2017ybb,Moult:2018jzp,Moult:2019vou,Ebert:2020sfi,Moult:2022xzt,Duhr:2022yyp} in the back-to-back limit which is referred as the Sudakov region.
At hadron colliders, the resummation of the transverse EEC (TEEC) in the back-to-back limit has been pushed to N$^3$LL accuracy~\cite{Gao:2019ojf,Gao:2023ivm}.
Regarding the EEEC, resummation becomes more subtle since there are multiple overlapping singular regions. To study the collinear limit, the simplest way is to project the full angular dependence into a one-dimensional space, namely the projected energy correlators~\cite{Chen:2020vvp}. The projected $N$-point correlator is defined as
\begin{equation}\label{eq:pecdef}
\frac{\df\sigma}{\df x_L}=\sum_n\sum_{1\leq i_1,\cdots i_N\leq n}\int\! \df \sigma \frac{\prod_{a=1}^N E_{i_a}}{Q^N}\delta(x_L-\text{max}\{x_{i_1,i_2},x_{i_1,i_3}, \cdots x_{i_{N-1}, i_N}\})\,,
\end{equation}
with $x_L$ the largest angle, and logarithms of $\ln(x_L)$ in the collinear limit $x_L\to0$ have been resummed to NNLL accuracy~\cite{Lee:2022ige,Chen:2023zlx,Lee:2024icn}. Very recently, this result was used by CMS collaboration to produce the most accurate $\alpha_s$ value from a jet substructure observable~\cite{CMS:2024mlf}. 

\begin{figure}[!htbp]
        \centering
        \subfloat[]{\label{fig:3_jet_config}
        \includegraphics[scale=1.2]{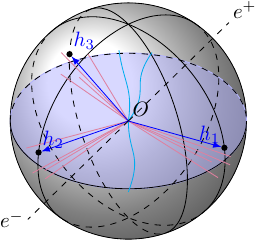}}
         \qquad \qquad
        \subfloat[]{\label{fig:tau_def}
        \includegraphics[scale=0.95]{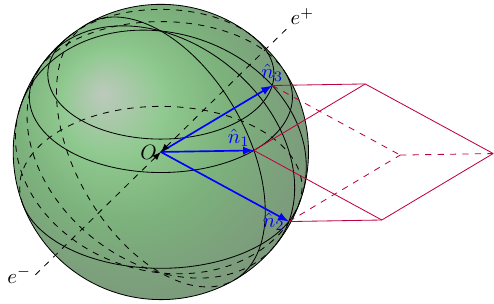}}
        \captionof{figure}{
         a) Configuration where $h_1,h_2,h_3$ are within three well-separated coplanar jets.
         b) Visualization for the $\tau_p$ projection of the EEEC. The three vectors $\hat{n}_1$, $\hat{n}_2$ and $\hat{n}_3$ represent the directions of three final-state hadrons $h_1$, $h_2$, $h_3$, and they form a parallelepiped, whose volume defines $\tau_p$.
         }
\end{figure}

In this paper, we initiate the study of coplanar limit of energy correlators in the trijet region, where the three particles assigned with the energy weight fall almost on the same plane (see Fig.~\ref{fig:3_jet_config}). To access this limit, we introduce a {\it volume projection} of the EEEC (referred as {\it $\tau_p$-projected EEEC}), defined as follows
\begin{equation}\label{eq:EEEC_tau_def}
         \frac{\df\sigma}{\df\tau_p}=\sum_{h_1,h_2,h_3}\int\! \df \sigma \, \frac{E_{h_1} E_{h_2}E_{h_3}}{Q^{3}}\, \delta\left(\tau_p-\tau_{h_1h_2h_3}\right), \quad \tau_{h_1h_2h_3}\equiv \left|\left(\hat{n}_1\times \hat{n}_2\right)\cdot \hat{n}_3\right|
        \,.
\end{equation} 
Here $h_1$, $h_2$, $h_3$ run over all the final-state hadrons, and $\tau_{h_1h_2h_3}$\footnote{Although we take the absolute value in the definition of $\tau_{h_1h_2h_3}$, it is also interesting to explore $\left(\hat{n}_1\times \hat{n}_2\right)\cdot \hat{n}_3$ without taking the absolute value, in order to study unconserved parity. In that case, the order of $\{h_1,h_2,h_3\}$ matters.
One may consider ordering $\{h_1,h_2,h_3\}$ by the order of their energies, or by their azimuthal angles along a preferred direction. We leave the exploration of these possibilities to future work.}
ranging from 0 to 1 is the volume of the parallelepiped formed by the unit vectors $\hat{n}_1$, $\hat{n}_2$ and $\hat{n}_3$ (as illustrated in \fig{tau_def}). Note that simply imposing $\tau_p\rightarrow 0$ is insufficient for infrared safety and selecting the interesting coplanar configurations.
This is because $\tau_{h_1h_2h_3}\to 0$ also includes contributions where two particles are collinear (within one jet) or back-to-back (in the two different jets of the dijet configuration), or three particles are collinear (within one jet). To resolve this problem, we use the $k_T$ (Durham) algorithm~\cite{Catani:1991hj} to extract the coplanar trijet limit. Such a procedure was also used in some trijet analysis of event shapes~\cite{Banfi:2001pb,Arpino:2019ozn}.
Explicitly, we impose a lower bound $y_\text{cut}$ for the 3-jet resolution variable $y_3$\footnote{$y_3$ is constructed as follows. Given a set of $n$ momenta, for each pair of momenta $(i,j)$, calculate the distance variable of the $k_T$ algorithm, $y_{ij}=2\min(E_i^2,E_j^2)(1-\cos\theta_{ij})/Q^2$. Find the smallest $y_{ij}$ and define it to be the $n$-jet resolution $y_n$. Combine such $i$ and $j$ into a single new pseudoparticle, with momentum $p=p_i+p_j$. Repeat this procedure until there are 3 (pseudo)particles left. Then $y_3$ is the 3-jet resolution of the remaining 3 (pseudo)particles.}~\cite{Catani:1991hj} to select coplanar three-jet events and modify the definition in \eq{EEEC_tau_def} to
\begin{equation}\label{eq:EEEC_cpln_def}
  \frac{\df\sigma}{\df\tau_p}\Bigg|_{\text{trijet}}= \sum\limits_{h_1\in J_1}\sum\limits_{h_2\in J_2}\sum\limits_{h_3\in J_3}\,\,\,
  \int\limits_{y_3>y_{\rm cut}} \df\sigma\, 
    \,\frac{E_{h_1}\, E_{h_2}\, E_{h_3}}{E_{J_1}\, E_{J_2}\,E_{J_3}}\, \delta\left(\tau_p-\tau_{h_1h_2h_3}\right)
    \,,
\end{equation}
where $J_1$, $J_2$, $J_3$ denote the three jets selected using the $k_T$ algorithm, and $E_{J_1}$, $E_{J_2}$, $E_{J_3}$ are the energies of the three jets. 
Below we will focus on this particular definition~\eqref{eq:EEEC_cpln_def}, referred to as {\it the coplanar EEEC}, but omit the subscript ``trijet''.

To further analyze different coplanar configurations, we can also characterize the coplanar limit using celestial coordinates $\{s,\phi_1,\phi_2\}$ introduced in Ref.~\cite{Yan:2022cye} and study the fully differential spectrum. As we will show in \sec{diffspectrum}, the scaling variable $s$ can access both the collinear limit ($s\to 0$) and coplanar limit ($s\to 1$), while values of $\phi_{1,2}$ fix specific shapes of the trijet. Since the factorization structure remains similar in either the $\tau_p$ definition or celestial definition, we will focus on the former case for the main part of this paper.

The fixed-order expansion of the coplanar EEEC in perturbation theory can be written in the following form:
\begin{equation}
   \frac{\df\sigma}{\df\tau_p}=\delta(\tau_p)\left[\left(\frac{\alpha_s}{4\pi}\right)A_1+\left(\frac{\alpha_s}{4\pi}\right)^2A_2+\cdots\right]+\left(\frac{\alpha_s}{4\pi}\right)^2 B(\tau_p)+\left(\frac{\alpha_s}{4\pi}\right)^3 C(\tau_p)+\cdots
    \,.
\end{equation}
Here, the delta function term arises from the three-particle configuration, where the three particles are exactly coplanar, and thus one has $\tau_p=0$. The leading order (LO) distribution with non-trivial $\tau_p$ dependence, $B(\tau_p)$, starts at $\mathcal{O}\left(\alpha_s^2\right)$ with four-particle final states, 
and can be computed using the analytic EEEC result~\cite{Yang:2022tgm}, or numerically using, e.g. \texttt{Event2}~\cite{Catani:1996jh,Catani:1996vz} and $\texttt{NLOJet++}$~\cite{Nagy:1998bb}.
The NLO contribution $C(\tau_p)$ including tree-level five-particle final states and one-loop four-particle final states,
can also be computed by $\texttt{NLOJet++}$.
Similar to the back-to-back limit of EEC, the coplanar EEEC receives double logarithmic series, i.e. 
$ \frac{\df\sigma}{\df\tau_p}\sim\sum_{k}\sum_{l\leq 2k-1}\alpha_s^k\left(\frac{\ln^l \tau_p}{\tau_p}\right)_{+}$, which requires resummation to all orders. In this work, we will develop a factorization formula for the coplanar EEEC in the framework of soft-collinear effective theory (SCET)~\cite{Bauer:2000ew, Bauer:2000yr, Bauer:2001ct, Bauer:2001yt} and perform the resummation to N${}^3$LL accuracy.

In the past few decades, extensive studies of various dijet event shapes, such as thrust~\cite{Gehrmann-DeRidder:2007vsv,Gehrmann-DeRidder:2007nzq,Becher:2008cf,Abbate:2010xh,Benitez-Rathgeb:2024ylc}, heavy-jet mass~\cite{Salam:2001bd,Chien:2010kc} as well as EEC, have been conducted at $e^+e^-$ colliders, establishing a comprehensive formalism for dijet resummation and non-perturbative corrections. 
However, calculations for trijet event shapes are scarce~\cite{Nagy:1997yn,Nagy:1998bb,Campbell:1998nn,Banfi:2000si,Banfi:2000ut,Banfi:2001pb,Banfi:2001sp,Larkoski:2018cke,Arpino:2019ozn}, 
and the study of the trijet configuration as power correction to traditional event shapes using modern approaches has only recently begun~\cite{Bhattacharya:2022dtm,Bhattacharya:2023qet,Luisoni:2020efy,Caola:2021kzt,Caola:2022vea,Nason:2023asn}.
The simplicity of coplanar EEEC and the existence of a non-perturbative definition for the energy correlator make this observable an ideal candidate for exploring both perturbative and non-perturbative QCD dynamics.
At future lepton colliders, such as CEPC~\cite{CEPCStudyGroup:2018rmc,CEPCStudyGroup:2018ghi,CEPCStudyGroup:2023quu}, ILC~\cite{Behnke:2013xla,ILC:2013jhg} and FCC-ee~\cite{TLEPDesignStudyWorkingGroup:2013myl}, the increased number of multi-jet events will enable precise measurements of the coplanar EEEC, allowing the determination of both Standard Model parameters (including $\alpha_s$) and the non-perturbative power corrections. Moreover, the formalism we develop for $e^+e^-\to\gamma^*\to$ trijet in this paper can be readily extended to other processes at the Large Hadron Collider (LHC). For example, our work has important implications for studying top quark decay, particularly given the recent interest in the potential of applying the EEEC to obtain precise top quark mass measurements~\cite{Holguin:2022epo,Holguin:2023bjf,Xiao:2024rol,Holguin:2024tkz}. As pointed out in Ref.~\cite{Holguin:2023bjf,pathak2024}, the Jacobian peak of the EEEC in top decay corresponds to exactly its coplanar limit in the top rest frame. Resumming the coplanar logarithms would lead to more accurate theoretical predictions for the EEEC of top decay and potentially improve the precision of the top mass measurements.

The outline of this paper is as follows. In Sec.~\ref{sec:factorization}, we discuss the trijet coplanar kinematics and derive the factorization theorem for \eq{EEEC_cpln_def}, focusing on the quark channel where the three jets are initiated by a quark, an antiquark, and a gluon.
We also discuss all the ingredients needed for resummation to N$^3$LL.
In \sec{resummation}, we compute the resummed distribution for $\tau_p$ to N$^3$LL and provide estimations for perturbative uncertainties.
In \sec{gluon_channel}, we consider the three-gluon jet case, whose contribution is $\cO(\alpha_s^2)$ suppressed.
In \sec{diffspectrum}, we focus on the fully differential distribution in terms of $\{s, \phi_1,\phi_2\}$, and study the $s\to1$ limit with $\phi_1$, $\phi_2$ fixed.
In \sec{analysis}, we investigate the non-perturbative corrections using the Monte Carlo program $\texttt{Pythia8}$, and derive a relation of the $\tau_p$-projected EEEC to $D$-parameter.

\section{Factorization in the coplanar limit}\label{sec:factorization}

In this section, we study the factorization theorem for the coplanar EEEC in the $\tau_p\to0$ limit. For $e^+e^-$ collisions, there are three partonic channels: 
(1). $e^+e^-\to \gamma^* /Z \to q\bar q g$; (2). ``light-by-glue'' process, $e^+e^-\to \gamma^* \to g g g$; (3). ``Z-by-glue'' process, $e^+e^-\to Z \to g g g$, as summarized in Ref.~\cite{Dixon:1997th}. Our paper will focus on the $\gamma^*$-mediated processes, including the $q\bar q g$ channel (two quark jets and one gluon jet) and $ggg$ channel (three gluon jets).
However, the analysis below can be applied to the other channels as well as other trijet processes. 

Similar to other trijet event shapes, the coplanar EEEC is expected to factorize into a hard function, three jet functions, and a soft function. Note that the hard function for the $ggg$ channel only starts at two-loop order, which is $\mathcal{O}(\alpha_s^2)$ suppressed compared to the $q\bar q g$ channel. For convenience, below we will use the $q\bar q g$ channel to derive the factorization theorem and the resummed formula, noting that these expressions can be extended to the $ggg$ channel by appropriate substitution of the corresponding ingredients.

\subsection{Kinematics}
\begin{figure}[!htbp]
  \centering
  \subfloat[]{\label{fig:coll_splt}
  \includegraphics[width=.35\linewidth]{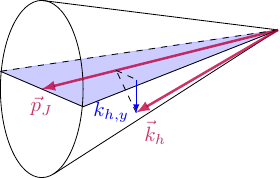}}
\qquad
 \subfloat[]{\label{fig:soft_rdia}\includegraphics[width=.48\linewidth]{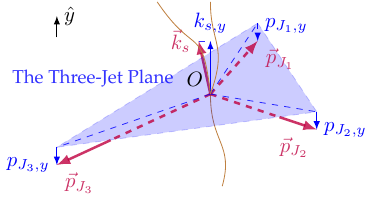}}
\qquad
\captionof{figure}{a) Collinear splitting within one jet. Hadron $h$ slightly deviates from the momentum direction of the jet.
b) The soft emissions accumulate to a total momentum  $\vec{k}_s$ , which results in a slight recoil of the three jets, pushing them slightly out of the trijet plane.
}
        \label{fig:jet&soft}
\end{figure}

Let us start by considering the partonic Born process $e^+\,e^-\,\to\gamma^*\to \,q(p_{q}) \,\bar q(p_{\bar q}) \, g(p_{g})$ where each of the final-state particles initiates a jet.
We will use $J_{1,2,3}$ and $\{q,\bar q,g\}$ interchangeably when discuss the kinematics: $p_{J_1}=p_q$, $p_{J_2}=p_{\bar q}$ and $p_{J_3}=p_g$. The coplanar three-jet limit defines a plane spanned by these three separated jets. Collinear splittings and soft emissions in the jets recoil the particles correlated by the EEEC observable slightly away from this plane. The kinematics of the factorization lies in a relation between the $\tau_p$ and the momentum perpendicular to the scattering plane, which we will denote as the $y$ component, as illustrated in \fig{jet&soft}. 

Now, consider three final-state hadrons with momenta  $k_{h_1}$,  $k_{h_2}$, and  $k_{h_3}$, originating from jets $J_1$, $J_2$ and $J_3$ respectively.
In addition to the recoil effect of the total soft momentum $k_{s,y}$, they also obtain transverse momenta from the momentum components off the scattering plane due to final-state collinear splittings. Explicitly, the soft recoil contribution to $\tau_{h_1h_2h_3}$ is proportional to $k_{s,y}$ times the area of the triangle (the blue triangle in \fig{soft_rdia}) formed by the three jets, $|\vec p_{J_1}\times\vec p_{J_2}|$, and a normalization factor $1/(E_{J_1}E_{J_2}E_{J_3})$ which reflects the fact that $\tau_{h_1h_2h_3}$ in Eq.~\eqref{eq:EEEC_tau_def} is defined using the unit vectors.
In the $\tau_p\to 0$ limit, we find the volume of parallelepiped to be
\begin{align}\label{eq:kinrel}
    \tau_{h_1h_2h_3} =\frac{\left|\vec p_{J_1}\times\vec p_{J_2}\right|}
                {E_{J_1}E_{J_2}E_{J_3}}
                \left|\frac{k_{h_1,y}}{z_{h_1}}+\frac{k_{h_2,y}}{z_{h_2}}+\frac{k_{h_3,y}}{z_{h_3}}-k_{s,y}\right|+\text{power corrections}\,,
\end{align}
where $z_{h_1}$ is the longitudinal momentum ratio of $k_{h_1}$ with respect to $p_{J_1}$, and similarly for $z_{h_2}$ and $z_{h_3}$. 
Collinear splittings of hadron $h$ deviated from the direction of $\vec p_J$, further adds a contribution of $k_{h,y}/z_h$.

For convenience, we define $\xi=\frac{\left|\vec p_{J_1}\times\vec p_{J_2}\right|Q}{E_{J_1}E_{J_2}E_{J_3}}$, which can be written in terms of jet energies
\begin{align}\label{eq:xi}
    \xi&=\frac{\left|\vec p_{J_1}\times\vec p_{J_2}\right|Q}{E_{J_1}E_{J_2}E_{J_3}}\nn\\
&=\frac{Q\sqrt{\left(E_{J_1}+E_{J_2}+E_{J_3}\right)\left(E_{J_2}+E_{J_3}-E_{J_1}\right)\left(E_{J_3}+E_{J_1}-E_{J_2}\right)\left(E_{J_1}+E_{J_2}-E_{J_3}\right)}}{2 E_{J_1} E_{J_2} E_{J_3}}\nn \\ 
&=\frac{4 \sqrt{uvw}}{(1-u)(1-v)(1-w)}\,.
\end{align}
In the second line, we have used Heron's formula for the area of a triangle with
sides $E_{J_1}$, $E_{J_2}$, $E_{J_3}$. In the third line, we introduce three dimensionless variables $u$, $v$ and $w$ via
\begin{equation}
u = \frac{s_{q\bar{q}}}{Q^2}\,, \quad v = \frac{s_{qg}}{Q^2} \,, \quad w = \frac{s_{\bar{q}g}}{Q^2} \,,\qquad \text{with }s_{ij}=(p_i+p_j)^2
\,,\end{equation}
satisfying $u+v+w = 1$ in the massless limit.

Putting everything together, at leading power, our measurement function takes the form of $\delta\left(\tau_p-\frac{\xi}{Q}\left|\frac{k_{h_1,y}}{z_{h_1}}+\frac{k_{h_2,y}}{z_{h_2}}+\frac{k_{h_3,y}}{z_{h_3}}-k_{s,y}\right| \right)$. Since the observable receives a contribution from perpendicular momenta, we will characterize the collinear splittings by the TMD fragmentation functions (TMDFFs) and the global soft emissions by a TMD trijet soft function. Now we are ready to write down the factorization theorem.

\subsection{Factorization theorem}

\begin{figure}[!htbp]
  \centering
\parbox{2.5in}{\includegraphics[width=\linewidth]{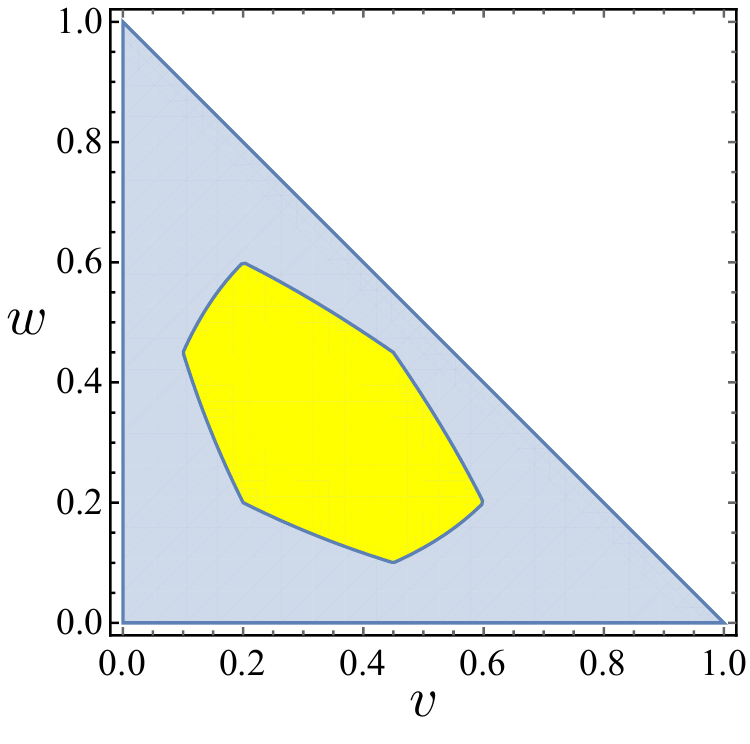}}
\captionof{figure}{The phase space $T$ for $y_3>y_{\rm cut}$ with $y_{\rm cut}=0.1$ is shown as the yellow region, plotted in $v$ and $w$ axises. The blue region is the total 3-body phase space.}
        \label{fig:ps_ycut}
\end{figure}

\begin{figure}
  \centering
\parbox{2.5in}{\includegraphics[width=\linewidth]{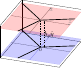}}
\captionof{figure}{The spatial structure of the coplanar EEEC soft function. Each set of Wilson lines lies on the trijet plane, and their relative displacement is perpendicular to the plane.}
        \label{fig:soft_wlsnl}
\end{figure}

Given the kinematic analysis, the coplanar EEEC cross-section is factorized as
\begin{align}\label{eq:full}
  &\frac{1}{\sigma_{\text{LO}}}\frac{\df\sigma_q}{\df\tau_p}  =\int_T \df v\,\df w~H_q(v,w,\mu)\sum_{h_1,h_2,h_3}\int\! \df  k_{h_1,y}\int\! \df  k_{h_2,y}\int\! \df  k_{h_3,y} \int\! \df  k_{s,y}\int\! \df  z_{h_1}\int\! \df  z_{h_2}\int\! \df  z_{h_3} \nn\\ 
  &\hspace{2cm}\times z_{h_1}z_{h_2}z_{h_3}\, F_{h_1/q}(k_{h_1,y}, z_{h_1},\mu,\nu)  F_{ h_2/\bar{q}}(k_{h_2,y}, z_{h_2},\mu,\nu) F_{h_3/g}(k_{h_3,y}, z_{h_3},\mu,\nu)  \nn\\
  &\hspace{2cm} \times S_q(k_{s,y},\mu,\nu)\,\delta\left(\tau_p-\xi\left|\frac{k_{h_1,y}}{z_{h_1}}+\frac{k_{h_2,y}}{z_{h_2}}+\frac{k_{h_3,y}}{z_{h_3}}-k_{s,y}\right|\right)+\text{power corrections}\,,
\end{align}
where $T$ is the domain of the $\df v$ and $\df w$ integral which is constrained by the phase-space cuts from the jet algorithm. For the case of $y_{\rm cut}=0.1$, $T$ is plotted in \fig{ps_ycut}. The subscript $q$ on the LHS denotes the $\gamma^*\to q\bar q g$ channel. We also adopt the normalization 
\begin{align}
\label{eq:sigmaLO}
    \sigma_{\text{LO}}=\frac{\alpha_s(Q)}{4\pi} C_F \, \sigma_0\,,
\end{align}
where $ \sigma_0$ is the Born cross-section for $e^+e^-\to q\bar q$. This ensures that the order counting of the coplanar EEEC matches that of traditional dijet event shape resummation.
$H_q$ and $S_q$ are the trijet hard function and the trijet TMD soft function respectively. The subscript $q$ here denotes the quark channel.
The hard function $H_q(v,w,\mu)$ is obtained by matching SCET to QCD. In fixed-order perturbation theory, it can be expanded as
$H_q(v,w,\mu) = \sum_{n=0} \left( \frac{\alpha_s(\mu)}{4 \pi} \right)^{n} H_q^{(n)}(v,w,\mu)$. The $\sigma_{\text{LO}}$ normalization in Eq.~\eqref{eq:full} implies that the hard function is normalized such that its leading order begins at $\mathcal{O}(\alpha_s^0)$. 
The soft function $S_q$ is defined as a vacuum expectation value of Wilson lines,
\begin{align}
  S_q(n_q, n_{\bar q}, n_g, b_y) = \mathrm{tr}
\langle 0 | \mathrm{T}[ \mathbf S_{n_q} \mathbf S_{n_{\bar q}} \mathbf S_{n_g} (0^\mu) ] \overline{\mathrm{T}} [ \mathbf S^\dagger_{n_q} \mathbf S^\dagger_{n_{\bar q}} \mathbf S_{n_g}^\dagger (0^+,0^-,0_x,b_y) ] | 0 \rangle \,,
\end{align}
where the soft Wilson line $\mathbf S_{n_i}$ is defined as
\begin{equation}
  \mathbf S_{n_i}(x)=\cP\exp\left(ig_s\int_{-\infty}^0\! \df s\, n_i\cdot A_s(x+sn_i)\mathbf T_i\right)\,.
\end{equation}
Here $b_y$ is the conjugate variable to $k_{s,y}$. We use the color space notation from Ref.~\cite{Catani:1996vz} and the trace is over the color indices. Notice that the Wilson lines are in either fundamental or adjoint representation, depending on $i$ being quark or gluon. 
The spatial structure of the Wilson lines is indicated in \fig{soft_wlsnl}.
Calculations of the soft function will be discussed in \sec{soft}.

In addition, $F_{h/i}$ with $i=q,\,\bar q,\,g$ are the TMD fragmentation functions (TMDFFs), which describe the collinear radiation of a final-state hadron $h$ from the splitting of the parton $i$, with energy fraction $z_h$ and $y$ component (perpendicular to the three-jet plane) of transverse momentum, $k_{h,y}$.
$F_{h/f}$ here are the TMDFFs in the parton frame, by which we mean the frame where the parton has zero transverse momentum. 
However, it is usually easier to give an operator definition of the TMDFFs in the hadron frame, where hadron $h$ has zero transverse momentum. We use $D$ to denote the hadron-frame TMDFFs, defined in position space as~\cite{Collins:2011zzd}
\begin{align}\label{eq:TMDFF_def_q}
D_{h/q}(\vec b_\perp,z_h)&
  = \frac{1}{2z_hN_c}
    \sum_{X} \int\!\frac{\df b^+}{4\pi}e^{i k_h^- b^+/(2z_h)}\,\tr_{\rm spin} \Mae{0}{\frac{\slashed\bn}{2}\chi_{n}(b)}{h,X}
    \Mae{h,X}{\bar\chi_{n}(0)}{0}
    \,,
    \\\label{eq:TMDFF_def_g}
D_{h/g}(\vec b_\perp,z_h)&
  = -\frac{k_h^-\, g_{\perp\mu\nu}}{(d-2)(N_c^2-1)z_h^2}
    \sum_{X} \int\!\frac{\df b^+}{4\pi}e^{i k_h^- b^+/(2z_h)}\, \Mae{0}{\cB_{n\perp}^\mu(b)}{h,X}
    \Mae{h,X}{\cB_{n\perp}^\nu(0)}{0}
    \,.
\end{align}
where $\chi_n$ and $\cB_n^\mu$ are in the SCET notation, denoting the gauge invariant $n$-collinear quark and gluon fields respectively.
Note that in \eq{TMDFF_def_g} we keep only the transverse polarization contribution of the gluon since the linear polarization does not contribute to the process considered here. The pair of fields are separated by a space-like distance $b^\mu=(b^+,0^-,\vec b_\perp)$, where $\vec b_\perp$ is the conjugate variable to the parton's transverse momentum relative to $\vec{n}$. A detailed discussion of the relation between the TMDFFs in the two frames can be found in Refs.~\cite{Collins:2011zzd,Luo:2019hmp,Luo:2019bmw}; Here we simply state the result,
\begin{align}
  F_{h/i}(\bt/z_h,z_h)=z_h^{2-2\epsilon}\, D_{h/i}(\bt,z_h)\,.
\end{align}

The operator product expansion (OPE) of the TMDFF onto the standard fragmentation functions is given in momentum space by
\begin{align}
\label{eq:TMDFF_matching}
F_{h/i}(\vec k_{h\perp},z_h) &=\sum \limits_j \int \frac{\df x_h}{x_h^3}d_{h/f} (x_h) \cJ_{fi}\left(\frac{\vec k_{h\perp}}{x_h},\frac{z_h}{x_h}\right)  
+ \text{power corrections}\,,
\end{align}
where $\cJ_{fi}$ are finite matching coefficients, 
and $d_{h/f}$ are the fragmentation functions. 
To convert these TMDFFs as well as their matching coefficients to the ones shown in \eq{full} as functions of the $y$-component momenta, one simply integrates out their $x$ components,
\begin{align}\label{eq:TMDFF_y}
    F_{h/i}(k_{h,y},z_h) 
    =\int\!\df k_{h,x}\, F_{h/i}(\vec k_{h\perp},z_h)
    \,,\qquad \text{and} \qquad
    \cJ_{fi}\left(k_y,\xi\right)=\int\!\df k_x\,\cJ_{fi}\left(\vec k_\perp,\xi\right)\,,
\end{align}
so that we have
\begin{align}
    \label{eq:TMDFF_matching_y}
    F_{h/i}(k_{h,y},z_h) 
    &=\sum \limits_f \int \frac{\df x_h}{x_h^2}d_{h/f} (x_h)\, \cJ_{fi}\left(\frac{k_{h,y}}{x_h},\frac{z_h}{x_h}\right)
    + \text{power corrections}\,.
\end{align}

Now let us plug the above equation into the factorization formula~\eqref{eq:full} and simplify the result on the right-hand side. Explicitly, we insert \eq{TMDFF_matching_y} into \eq{full}, and change the variables from $\{z_{h_1}, x_{h_1}\}$ to $\{\zeta_{f_1}, x_{h_1}\}$
\begin{equation}
\zeta_{f_1}=\frac{z_{h_1}}{x_{h_1}},\;\;\;\df z_{h_1} \df x_{h_1}=x_{h_1} \df\zeta_{f_1} \df x_{h_1}\,,
\end{equation}
and similarly from $\{z_{h_2}, x_{f_2}\}$ to $\{\zeta_{f_2}, x_{h_2}\}$ and from $\{z_{h_3}, x_{h_3}\}$ to $\{\zeta_{f_3}, x_{h_3}\}$.
We then find
\begin{align}
  &\frac{1}{\sigma_{\text{LO}}}\frac{\df\sigma_q}{\df\tau_p}  =\int_D \df v\df w~H_q(v,w,\mu)\sum_{f_1,f_2,f_3}\int\! \df  k_{f_1,y}\int\! \df  k_{f_2,y}\int\! \df  k_{f_3,y} \int\! \df  k_{s,y}\int\! \df  \zeta_{f_1}\int\! \df  \zeta_{f_2}\int\! \df  \zeta_{f_3} ~ \zeta_{f_1}\zeta_{f_2}\zeta_{f_3}  \nonumber\\
  &\times S(k_{s,y},\mu,\nu)\,\delta\left(\tau_p-\xi\left|\frac{k_{f_1,y}}{\zeta_{f_1}}+\frac{k_{f_2,y}}{\zeta_{f_2}}+\frac{k_{f_3,y}}{\zeta_{f_3}}-k_{s,y}\right|\right) \left[\sum\limits_{h_1} \int\! \df  x_{h_1}\, x_{h_1}\,d_{h_1/f_1}(x_{h_1},\mu) \right]\cJ_{f_1 q}\left(k_{f_1,y},\zeta_{f_1}\right) \nn\\
  &\times\left[\sum\limits_{h_2} \int\! \df  x_{h_2}\, x_{h_2}\,d_{h_2/f_2}(x_{h_2},\mu) \right] \cJ_{f_2\bar{q}}\left(k_{f_2,y},\zeta_{f_2}\right)\, \left[\sum\limits_{h_3} \int\! \df  x_{h_3}\, x_{h_3}\,d_{h_3/f_3}(x_{h_3},\mu) \right]\cJ_{f_3 g}\left( k_{f_3,y},\zeta_{f_3}\right)
  \nn\\&
  +\text{power corrections}
   \,,
\end{align}
with $k_{f_1,y}=\frac{k_{h_1,y}}{x_{h_1}}$, $k_{f_2,y}=\frac{k_{h_2,y}}{x_{h_2}}$, $k_{f_3,y}=\frac{k_{h_3,y}}{x_{h_3}}$. The integrals inside the square brackets are simply unity due to the momentum-conservation sum rule
\begin{align}
    \sum\limits_h \int\! \df  x~ x~ d_{h/i}(x) =1\,.
\end{align}

To further simplify the factorization formula, we define the position-space matching coefficients and the soft function
\begin{align}
   \cJ_{fi}\left(b_y\right) &=\int\! \df  k_y~ e^{i b_y k_y}~\cJ_{fi}\left(k_y,\zeta\right)\,, \\
   S(b_y)&=\int\! \df  k_{s,y}~ e^{i b_y k_{s,y}}~S\left(k_{s,y}\right)\,,
\end{align}
as well as the jet function
\begin{align}\label{eq:J_NP}
J_i(b_y) = \sum\limits_f  \int _0^1 \! \df\zeta~ \zeta~ \cJ_{fi}\left(\frac{b_y}{\zeta},\zeta\right),\quad i,f\in \{q,\bar q,g\}\,,
\end{align}
and apply the Fourier representation of the measurement function
 \begin{align}
 \label{eq:jet_def}
 \delta\left(\tau_p-\xi\left|\frac{k_{f_1,y}}{\zeta_{f_1}}+\frac{k_{f_2,y}}{\zeta_{f_2}}+\frac{k_{f_3,y}}{\zeta_{f_3}}-k_{s,y}\right|\right)
  &= \int_{-\infty}^\infty \frac{\df b_y}{2 \pi \xi} ~2\cos\left( b_y \tau_p / \xi\right) \nonumber \\
  &\quad \times \exp\left[ i b_y\left(\frac{k_{f_1,y}}{\zeta_{f_1}}+\frac{k_{f_2,y}}{\zeta_{f_2}}+\frac{k_{f_3,y}}{\zeta_{f_3}}-k_{s,y}\right)\right] \,.
 \end{align}
Then we arrive at the final factorization formula for the EEEC in the coplanar limit
\begin{align}\label{eq:factexp}
\frac{1}{\sigma_{\text{LO}}}\frac{\df\sigma_q}{\df\tau_p}
  =\,&\int_T\! \df v \df w H_q(v,w,\mu)\int_{-\infty}^{\infty}\!\frac{\df b_y}{2\pi\xi}\, 2\cos\left( \frac{b_y \tau_p} { \xi}\right)S(b_y,\mu,\nu)  J_q\left(b_y,\mu,\nu\right)
J_{\bar{q}}\left(b_y,\mu,\nu\right)J_g\left(b_y,\mu,\nu\right)\nn\\
&+\text{power corrections}
\,.
\end{align}

\subsection{Ingredients}

In this subsection, we will discuss all the ingredients appearing in the final factorized formula \eqref{eq:factexp}, as well as the UV and rapidity renormalization group equations (RGEs) they satisfy. In particular, we will see that up to two-loop accuracy, our trijet soft function factorizes into a product of three dipole TMD soft functions, which are known to three loops analytically~\cite{Li:2016ctv,Vladimirov:2016dll}. The simplicity of the soft function is one main reason that allows us to reach the N$^3$LL accuracy for the coplanar EEEC.

\subsubsection{Hard Function}
\label{sec:hard}
The hard function can be obtained by matching the QCD current into the corresponding SCET current operator. At leading order, the hard function is simply proportional to the tree-level process $e^+e^-\rightarrow q\bar q g$:
\begin{equation}
\label{eq:hard0qqg}
  H_q^\zero (v,w,\mu) =2 \frac{(1-v)^2+(1-w)^2}{vw} \,,
\end{equation}
where subscript $q$ denotes the quark channel.
At NLO and beyond, since all virtual graphs in SCET are scaleless in dimensional regularization, we can extract the trijet hard function from the virtual contribution in squared QCD matrix elements $e^+e^-\rightarrow 3 \text{ partons}$. Such a hard function is the same as direct photon production or 3-jetness by crossing. We begin our extraction of the hard function at NNLO accuracy by examining the form factor results for $e^+ e^- \to q \bar{q} g$ up to NNLO~\cite{Garland:2001tf,Garland:2002ak,Gehrmann:2023zpz}. For our purposes, we utilize the UV-renormalized and IR-unsubtracted results from the decay region as presented in Ref.~\cite{Gehrmann:2023zpz}. It is worth noting that the results in Ref.~\cite{Gehrmann:2023zpz} are presented with $\mu^2 = Q^2$. Given that scale dependence can arise from both UV and IR origins, we opt to reconstruct the scale by first outlining a general form for the UV-unrenormalized and IR-unsubtracted form factors:
\begin{align}
\label{eq:bareM}
    \bigl\lvert \mathcal{M}^{\text{bare}}_{q\bar{q}g} \bigr\rangle =g_s^{\text{bare}} \bigg\{ \bigl\lvert \mathcal{M}^{(0)}_{q\bar{q}g} \bigr\rangle+ \frac{\alpha_s^\text{bare}}{4 \pi} \left(\frac{Q^2}{\mu^2}\right)^{- \epsilon}  \bigl\lvert \mathcal{M}^{(1)}_{q\bar{q}g} \bigr \rangle +  \left(\frac{\alpha_s^\text{bare}}{4 \pi}\right)^2 \left(\frac{Q^2}{\mu^2}\right)^{- 2\epsilon}  \bigl\lvert \mathcal{M}^{(2)}_{q\bar{q}g} \bigr\rangle + \mathcal{O}(\alpha_s^3) \bigg\}\,.
\end{align}
It is important to observe that $\bigl\lvert \mathcal{M}^{(i)}_{q\bar{q}g} \bigr\rangle$ for $i=0,1,2$ are independent of the scale $\mu$; all scale dependence is derived from the prefactor $\left(\frac{Q^2}{\mu^2}\right)^{-\epsilon}$. Following this, we conduct the strong coupling renormalization by implementing these substitutions:
\begin{align}
\alpha_s^\text{bare} \to \alpha_s Z_{\alpha_s}\,, \nonumber \
g_s^{\text{bare}} = g_s \sqrt{Z_{\alpha_s}}\,,
\end{align}
where
\begin{align}
Z_{\alpha_s} = 1- \frac{\beta_0}{\epsilon} \frac{\alpha_s}{4\pi} - \left( \frac{\beta_0^2}{\epsilon^2} - \frac{\beta_1}{2 \epsilon} \right)\left(\frac{\alpha_s}{4\pi}\right)^2 + \mathcal{O}(\alpha_s^3)\,,
\end{align}
Here, $\beta_i$ represents the QCD beta function, which can be found in \app{betaRG}. This UV renormalization procedure yields UV-finite results, allowing us to determine $\bigl\lvert \mathcal{M}^{(i)}_{q\bar{q}g} \bigr\rangle$ for $i=0,1,2$ by comparing our results with those in Ref.~\cite{Gehrmann:2023zpz}. This method enables us to recover the scale dependence for the form factor.
Next, we address the IR subtraction for the UV-finite form factors. The IR singularities of massless scattering amplitudes are well-understood~\cite{Catani:1998bh,Becher:2009cu,Becher:2009qa,Gardi:2009qi} and are governed by a multiplicative renormalization factor $\mathbf{Z}$. In our context, this relationship is expressed as:
\begin{align}
\label{eq:finiteR}
\bigl \lvert \mathcal{M}^{\text{finite}}{q\bar{q}g} \bigr \rangle = \mathbf{Z}^{-1}  \bigl \lvert \mathcal{M}^{\text{}}{q\bar{q}g} \bigr \rangle \,,
\end{align}
In this equation, $\bigl \lvert \mathcal{M}^{\text{}}{q\bar{q}g} \bigr \rangle$ represents the UV-renormalized but IR-unsubtracted amplitude, while $\mathbf{Z}^{-1}$ serves to minimally subtract the IR divergences in $\bigl \lvert \mathcal{M}^{\text{}}{q\bar{q}g} \bigr \rangle$, ensuring that the left-hand side of the equation remains finite. Finally, we extract the hard function with proper normalization:
\begin{align}
H_q(v,w,\mu) = \frac{1}{ C_F\,\alpha_s/(4\pi)} \bigl \langle \mathcal{M}^{\text{finite}}{q\bar{q}g}  \bigl \lvert \mathcal{M}^{\text{finite}}{q\bar{q}g} \bigr \rangle\,.
\end{align}
In the ancillary file, we provide the expressions for the hard function up to two loops.

The extracted hard function satisfies the following RG equation,
\begin{align}\label{eq:hard_RGE}
  \frac{\df \ln H_q(v,w,\mu)}{\df\ln\mu^2} =&-\gamma_{\rm cusp}(\alpha_s)\sum_{1\leq i<j\leq3}\bT_i\cdot \bT_j \ln\frac{s_{ij}}{\mu^2}
  + \sum_{i=1}^3\gamma_i(\alpha_s) 
  \\& \nn
  +\gamma_{\rm quad}(\alpha_s) \, f_{abe}f_{cde} \, \sum_{i=1}^3\sum_{\substack{{1\leq j<k\leq 3}\\ j,k\neq i}}\bigl\{\bT_i^a,  \bT_i^d\bigr\}   \bT_j^b \bT_k^c  
  +\cO(\alpha_s^4)\,,
\end{align}
where $\bT_i$'s are Catani-Seymour's color insertion operator~\cite{Catani:1996vz}. 
The $\bT_i\cdot\bT_j$ term in \eq{hard_RGE} is the dipole contribution. 
The second line is the quadrupole color contribution (involving four color operators)
which starts at three loops and was calculated in \Refs{Almelid:2015jia,Almelid:2017qju}, which is a constant without any kinematic dependence,\footnote{
For processes that involve four or more partons, the quadrupole contribution includes terms involving four lightlike directions and has nontrivial dependence on kinematics.
The nontrivial kinematic dependence comes through a cross ratio of four lightlike vectors.
For the case here, the quadrupole contribution is a constant since it is impossible to construct a nontrivial scaling invariant from three lightlike vectors.}
\begin{align}
    \gamma_{\rm quad}=\left(\frac{\alpha_s}{4\pi}\right)^3 16\, (\zeta_5 + 2\zeta_2\,\zeta_3)+\cO(\alpha_s^4)
    \,.
\end{align}

For the $q\bar q g$ color singlet state, we can replace $\bT_i \cdot \bT_j$ in \eq{hard_RGE} with the Casimir factors by making use of the overall color conservation $\bT_q + \bT_{\bar q} + \bT_g = 0$,
\begin{subequations}\label{eq:casm}
    \begin{align}\label{eq:casm1}
  \bT_q \cdot \bT_g =&\ \frac{1}{2} ( \bT_{\bar q}^2 - \bT_q^2 - \bT_g^2) = - \frac{C_A}{2} \,,
\\\label{eq:casm2}
  \bT_{\bar q} \cdot \bT_g =&\ \frac{1}{2} ( \bT_{q}^2 - \bT_{\bar q}^2 - \bT_g^2) = - \frac{C_A}{2} \,,
\\\label{eq:casm3}
  \bT_q \cdot \bT_{\bar q} =&\ \frac{1}{2} ( \bT_{g}^2 - \bT_q^2 - \bT_{\bar q}^2) =  \frac{C_A - 2 C_F}{2} \,.
\end{align}
\end{subequations}
And the color quadrupole color factor can be calculated to be
\begin{align}\label{eq:quad_color}
    f_{abe}f_{cde} \, \sum_{i=1}^3\sum_{\substack{{1\leq j<k\leq 3}\\ j,k\neq i}}\bigl\{\bT_i^a,  \bT_i^d\bigr\}   \bT_j^b \bT_k^c
    \,\bigl\lvert q\bar q g\bigr\rangle\,=\,  \bigl\lvert q\bar q g\bigr\rangle\,\frac32 C_A
    \,,
\end{align}
by noticing that the color state $\bigl\lvert q^\alpha\bar q^\beta g^a\bigr\rangle$ has only one color structure which is proportional to $T_{\alpha\beta}^a$, where $\alpha,\,\beta,\,a$ are color indices for $q,\,\bar q,\,g$ respectively.

It is convenient to absorb the kinematic dependence in the cusp piece into the non-cusp piece. This leads to the final RGE for the hard function
\begin{equation}
\label{eq:hardRGE_cuspNcusp}
    \frac{\df\ln H_q(v,w,\mu)}{\df\ln\mu^2} = \frac{C_A + 2 C_F}{2} \gamma_{\rm cusp}(\alpha_s) \ln \frac{Q^2}{\mu^2}  + \gamma_{h,q}(v,w,\alpha_s) \,,
\end{equation}
where we define the non-cusp piece to be
\begin{align}\nn
  \gamma_{h,q}(v,w,\alpha_s) =\, &\gamma_{\rm cusp}(\alpha_s) \left( \frac{C_A}{2} \ln v + \frac{C_A}{2} \ln w + \frac{2 C_F - C_A}{2} \ln(1-v-w) \right)\nn\\
  &+ 2 \gamma_q(\alpha_s) + \gamma_g(\alpha_s)
 \,+\frac32\,C_A\gamma_{\rm quad}[\alpha_s]
 \,.
\end{align}
The solution to the RG equation is formally given by
\begin{align}\label{eq:hard_sol}
  H_q(v,w,\mu) = H_q(v,w,\mu_h) \exp \left[ \int_{\mu_h^2}^{\mu^2} \frac{\df\bar{\mu}^2}{\bar{\mu}^2}  \left(\frac{C_A + 2 C_F}{2} \gamma_{\rm cusp}(\alpha_s(\bar\mu)) \ln \frac{Q^2}{\bar \mu^2} + \gamma_{h,q}(v,w,\alpha_s(\bar\mu))  \right)\right] \,.
\end{align}

For convenience, we define the following RG notations:
\begin{align}\label{eq:RGkernel_def}
    S_{\Gamma}(\nu,\mu)&=-\int_{\alpha_s(\nu)}^{\alpha_s(\mu)}\df\alpha\frac{\gamma_{\text{cusp}}(\alpha)}{\beta(\alpha)} \int_{\alpha_s(\nu)}^\alpha \frac{\df\alpha^\prime}{\beta(\alpha^\prime)},\nn\\
    A_{\Gamma}(\nu,\mu)&=-\int_{\alpha_s(\nu)}^{\alpha_s(\mu)}\df\alpha\frac{\gamma_{\text{cusp}}(\alpha)}{\beta(\alpha)},\nn\\
    A_{\gamma_{j}}(\nu,\mu)&=-\int_{\alpha_s(\nu)}^{\alpha_s(\mu)}\df\alpha\frac{\gamma_{j}(\alpha)}{\beta(\alpha)}\,,
\end{align}
Then \eq{hard_sol} can be rewritten as
\begin{equation}
    H_q(v,w,\mu)=H_q(v,w,\mu_h)\exp\left[2(C_A+2C_F)S_{\Gamma}(\mu_h,\mu)-2A_{\gamma_{h,q}}(\mu_h,\mu)\right]\left(\frac{Q^2}{\mu_h^2}\right)^{-(C_A+2C_F)A_{\Gamma}(\mu_h,\mu)}\,.
\end{equation}
The explicit expressions of the running coupling and these RG kernels are given in~\app{betaRG}.

\subsubsection{Jet function}

The jet functions needed for the coplanar EEEC defined in \eq{jet_def} are the same as the jet functions in both the back-to-back limit of the EEC~\cite{Moult:2018jzp}  and the TEEC~\cite{Gao:2019ojf,Gao:2023ivm}. The fixed-order calculation has been pushed to N${}^3$LO~\cite{Luo:2020epw,Ebert:2020qef}, and we quote the one-loop result for illustration,
\begin{align}
\label{eq:oneloopjet}
  J_{q}(b_y,\mu,\nu)= J_{\bar q}(b_y,\mu,\nu)= \,&  1+\left(\frac{\alpha_s}{4\pi}\right)  (-2C_F L_bL_{\omega}+3C_F L_b+c_q^1)+\cO\left(\alpha_s^2\right)\,,
\nn\\
  J_{g}(b_y,\mu,\nu)=  \, & 1+\left(\frac{\alpha_s}{4\pi}\right)\left[-2C_A L_b L_{\omega}+\beta_0 L_b+c^g_1\right]+\cO\left(\alpha_s^2\right)\,.
\end{align}
Here $L_b = \ln (b_y^2 \mu^2/b_0^2)$ and $L_{\omega}=\ln\omega^2/\nu^2$, where $\omega=\bar n_i\cdot p_i$ is the large momentum of the jet $i$ ($\bar n_i$ here is the auxiliary lightcone vector opposite to the $n_i$ direction satisfying $n_i\cdot\bar n_i=2$). Explicitly, for the three jets $p_q$, $p_{\bar q}$ and $p_g$, we have 
\begin{equation}
  \label{eq:jete}
  \omega_q
  = Q (1-w) \,, \qquad \omega_{\bar q}
  = Q (1-v) \,, \qquad 
  \omega_g
  = Q (v+w) \,.
\end{equation}
The one-loop jet constants are 
\begin{equation}
  c_q^1=C_F(4-8\zeta_2)\,,\quad c_g^1=\left(\frac{65}{18}-8\zeta_2\right)C_A-\frac{5}{18} n_f\,.
\end{equation}

The jet functions satisfy the following UV and RG equations,
\begin{align}
  \label{eq:jetRG}
  \frac{\df  \ln J_i}{\df\ln \mu^2} &=  - \frac{1}{2} c_i \gamma_{\rm cusp} \ln \frac{\omega_i^2}{\nu^2} + \gamma_{J,i} \,,\\
  \label{eq:jetnuRG}
  \frac{\df \ln J_i}{\df\ln \nu^2} &= \frac{1}{2}c_i\left(\int_{b_0^2/b_y^2}^{\mu^2} \frac{\df \bar{\mu}^2}{\bar{\mu}^2}  \gamma_{\rm cusp}[\alpha_s(\bar{\mu})] - \gamma_r[\alpha_s(b_0/b_y)] \right) \,.
\end{align}
with $i=q,g$ for quark/gluon jet function, $\gamma_{J,i}$ the jet anomalous dimensions and $\gamma_r$ the rapidity anomalous dimension. These results are summarized in \app{anomalous_dimensions}.
Solving these equations gives the following solution,
\begin{align}\label{eq:jet_resum}
    J_i(\mu,\nu)=J_i(\mu_j,\nu_j)\left(\frac{\nu^2}{\nu_j^2}\right)^{-c_i A_\Gamma\left(\frac{b_0}{b_y}, \mu_j\right)-\frac{1}{2}c_i \gamma_r\left[\alpha_s\left(\frac{b_0}{b_y}\right)\right]}\left(\frac{\omega_i^2}{\nu^2}\right)^{c_i A_\Gamma(\mu_j,\mu)} e^{-2A_{\gamma_{J,i}}(\mu_j,\mu)}
    \,.
\end{align}

\subsubsection{Soft Function}\label{sec:soft}

The trijet soft function in the coplanar EEEC is the same as the TEEC soft function with three Wilson lines~\cite{Gao:2019ojf,Gao:2023ivm} (detailed discussion for it can be found in Ref.~\cite{Gao:2023ivm}). 
For the purpose of this paper, to achieve the N$^3$LL accuracy, we need the two-loop constants and the three-loop anomalous dimensions.
Thanks to the non-abelian exponentiation theorem~\cite{Gatheral:1983cz,Frenkel:1984pz}, the soft function can be written as the exponential of web diagrams.
Up to two loops, this soft function has only dipole contributions,\footnote{For the case of four-Wilson-line TEEC soft function, there is a tripole contribution (involving three Wilson lines) at two loops which is purely imaginary. As shown in Ref.~\cite{Gao:2023ivm}, such contribution vanishes for the three-Wilson-line soft function here due to color conservation and antisymmetry of $f^{abc}$.}
    \begin{align}
    \label{eq:softFunctionForm}
  S_q=\exp\biggl[-\sum_{1\leq i<j\leq 3} \bT_i \cdot \bT_j\, S_{ij}+\cO(\alpha_s^3)
    \biggr]\,.
\end{align}
where we have used the notation $\{1,2,3\}$ and $\{q,\bar q,g\}$ interchangeably, and $S_{ij}$ is the dipole contribution with lightcone directions $n_i$ and $n_j$.
$S_{ij}$ here can be simply written as
\begin{align}
    S_{ij}=S_{\perp}\,\Bigl(L_b,L_\nu+\ln\frac{n_i\cdot n_j}{2}\Bigr)
    \,,
\end{align}
where $L_\nu=\ln\nu^2b_y^2/b_0^2$, and $S_\perp(L_b,L_\nu)$ is the TMD soft function for color singlet production at hadron colliders (defined as the vacuum matrix element of back-to-back soft Wilson lines),
which can be found up to three loops in \cite{Li:2016ctv}.
In other words, going from the back-to-back dipole soft function to general $n_i$, $n_j$ lightcone directions, amounts to shift the rapidity logarithm $L_\nu$ by a boost factor $\ln(n_i\cdot n_j)/2$.

Starting from three loops, there are also contributions involving three Wilson lines, and it would be interesting to calculate them using the soft current in~\cite{Catani:2019nqv}. The scale-dependent part of this three-Wilson-line contribution can be predicted by the quadrupole term in \eq{RG_soft_UV} below, while the scale-independent part is beyond the working order of this paper.

The RG equations for the soft function are
\begin{align}\label{eq:RG_soft_UV}
  \frac{\df\ln S_q(n_q, n_{\bar q}, n_g, b_y, \mu, \nu)}{\df\ln \mu^2} 
=& -\sum_{1\leq i<j\leq 3} \bT_i \cdot \bT_j \left( \gamma_{\rm cusp} (\alpha_s) \ln \frac{\mu^2}{\nu^2 (n_i \cdot n_j)/2} - \gamma_s(\alpha_s) \right)
  \nn  \\
  & - f_{abe}f_{cde} \, \sum_{i=1}^3\sum_{\substack{{1\leq j<k\leq 3}\\ j,k\neq i}}\bigl\{\bT_i^a,  \bT_i^d\bigr\}   \bT_j^b \bT_k^c  \,\gamma_{\rm quad}(\alpha_s)
  +\cO(\alpha_s^4)
  \,,
\end{align}
\begin{equation}\label{eq:RG_soft_rap}
    \frac{\df\ln  S_q(n_q, n_{\bar q}, n_g, b_y, \mu, \nu)}{\df\ln \nu^2} = - \sum_{1\leq i<j\leq 3} \bT_i \cdot \bT_{j} \left( \int_{\mu^2}^{b_0^2/b_y^2} \frac{\df\bar \mu^2}{\bar \mu^2} \gamma_{\rm cusp}(\alpha_s[\bar \mu]) + \gamma_r (\alpha_s[b_0/b_y])  \right) \,.
\end{equation}
The first line in \eq{RG_soft_UV} is the dipole contribution and the second line is the quadrupole contribution. Similar structure also appeared for the hard function in \eq{hard_RGE}.
As discussed in the hard function section, note that
$\gamma_{\rm quad}$ does not have kinematic dependence on $v$ and $w$,
since it is impossible to construct a scaling invariant from three light-like vectors.

Similar to \sec{hard}, here we can also replace $\bT_i \cdot \bT_j$ in \eqs{RG_soft_UV}{RG_soft_rap} by the overall Casimir factors using \eq{casm}, and replace the quadrupole color factor with $\frac{3}{2}C_A$ using \eq{quad_color}.
We then write $n_i\cdot n_j$ in terms of kinematic variables $v$ and $w$,
\begin{align}
  \label{eq:nij}
 \frac{n_q \cdot n_{\bar q}}{2} = \, \frac{(1-v-w)}{(1-v) (1-w)} \,,\quad
 \frac{n_q \cdot n_g}{2} = \, \frac{v}{(v+w) (1-w)} \,,\quad
 \frac{n_{\bar q} \cdot n_g}{2} = \, \frac{w}{(1-v) (v+w)} \,.
\end{align}
This leads to the final RG equations for the trijet soft function
\begin{align}
  \label{eq:softRGE}
  \frac{\df\ln S_q}{\df\ln \mu^2} = &\, \left[\frac{2 C_F + C_A }{2} \left(\gamma_{\rm cusp}[\alpha_s] \ln \frac{\mu^2}{\nu^2} - \gamma_s[\alpha_s] \right) + C_F \gamma_{\rm cusp}[\alpha_s]\ln \frac{(1-v) (1-w)}{1 - v - w}
\right.\nn\\
&\, \left. + \frac{C_A}{2} \gamma_{\rm cusp}[\alpha_s] \ln \frac{(v+w)^2 (1 - v - w)}{vw}
\,-\frac32\,C_A\gamma_{\rm quad}[\alpha_s]
\right] \,,\\
  \label{eq:rapidityRGE}
  \frac{\df\ln S_q}{\df\ln \nu^2} =& \frac{2 C_F + C_A}{2}  \left( \int_{\mu^2}^{b_0^2/b_y^2} \frac{\df\bar \mu^2}{\bar \mu^2} \gamma_{\rm cusp}(\alpha_s[\bar \mu]) + \gamma_r (\alpha_s[b_0/b_y])  \right)  \,.
\end{align}
A solution is then given by
\begin{align}
  \label{eq:soft_sol}
  &S_q(\mu,\nu)  =S_q(\mu_s, \nu_s) \left(\frac{\nu^2}{\nu_s^2}\right)^{-(2C_F+C_A)A_\Gamma(\mu_s,b_0/b_y)+\frac{2C_F+C_A}{2}\gamma_r\left[\alpha_s\left(b_0/b_y\right)\right]} \left(\frac{\mu_s^2}{\nu^2}\right)^{-(2C_F+C_A)A_\Gamma(\mu_s,\mu)}\nn\\
  &\qquad\times \exp\left[-2(2C_F+C_A)S_\Gamma(\mu_s,\mu)+(2C_F+C_A)A_{\gamma_s}(\mu_s,\mu)
  +3C_A\, A_{\gamma_{\rm quad}}(\mu_s,\mu)
  \right]\nn\\
  &\qquad\times \exp\left[-2C_F\ln\frac{(1-v)(1-w)}{1-v-w}A_\Gamma(\mu_s,\mu)-C_A\ln\frac{(v+w)^2(1-v-w)}{v w}A_\Gamma(\mu_s,\mu)\right]
  \,.
\end{align}  

\subsection{Resummed formula}
Putting everything together in Eq.~\eqref{eq:factexp}, we obtain the resummed expression for the coplanar EEEC. Using the following identities
\begin{align}
S_\Gamma(\mu_1,\mu_2)+S_\Gamma(\mu_2,\mu_3)&=S_\Gamma(\mu_1,\mu_3)+\ln\frac{\mu_1}{\mu_2}A_\Gamma(\mu_2,\mu_3)\,,\nn\\    A_\Gamma(\mu_1,\mu_2)+A_\Gamma(\mu_2,\mu_3)&=A_\Gamma(\mu_1,\mu_3)\,,
\end{align}
and the relations between anomalous dimensions
\begin{equation}
    2\gamma_{h,q}+4\gamma_{J,q}+2\gamma_{J,g}-(2C_F+C_A)\gamma_s \,-3\,C_A\gamma_{\rm quad}-\ln(1-v-w)(2C_F-C_A)\gamma_{\text{cusp}}-\ln(vw)C_A\gamma_{\text{cusp}}=0\,,
\end{equation}
we can simplify it into the following form:
\begin{align}\label{eq:Finalresum}
    \frac{1}{\sigma_{\text{LO}}} \frac{\df\sigma_q}{\df\tau_p} &= \int_T\! \df v\, \df w\, H_q(v,w,\mu_h)\int_{-\infty}^{\infty}\frac{\df b_y}{2 \pi \xi} \, 2\cos\left(\frac{b_y\tau_p}{\xi}\right)\nn\\
    &\times S_q(\mu_s, \nu_s) J_q (\mu_j, \nu_j) J_{\bar q} (\mu_j, \nu_j) J_g (\mu_j, \nu_j)\times W_q(\mu_h, \mu_s, \nu_s,\mu_j, \nu_j)\,,
\end{align}
with the RG kernel
\begin{align}\label{eq:Wkernel}
    W_q(\mu_h, \mu_s, \nu_s,\mu_j, \nu_j) &= \exp\left[2(2C_F+C_A)(S_\Gamma(\mu_h,\mu_j)-S_\Gamma(\mu_s,\mu_j))\right]\nn\\
    &\times \exp\left[-2A_{\gamma_{h,q}}(\mu_h,\mu_s)+2A_{\gamma_{J,q}}(\mu_s,\mu_j)+2A_{\gamma_{J,q}}(\mu_s,\mu_j)+2A_{\gamma_{J,g}}(\mu_s,\mu_j)\right]\nn\\
    &\times\left(\frac{Q^2}{\mu_h^2}\right)^{(2C_F+C_A)A_\Gamma(\mu_j,\mu_h)}\times \left(\frac{\nu_j^2}{\nu_s^2}\right)^{\frac{2C_F+C_A}{2}\gamma_r\left[\alpha_s\left(b_0/b_y\right)\right]-(2C_F+C_A)A_\Gamma(\mu_j,b_0/b_y)}\nn\\
    &\times \left(\frac{\mu_s^2}{\nu_s^2}\right)^{(2C_F+C_A)A_\Gamma(\mu_j,\mu_s)}\times \left[(1-w)(1-v)\right]^{2C_F A_\Gamma(\mu_j,\mu_s)}(v+w)^{2C_A A_\Gamma(\mu_j,\mu_s)}\,.
\end{align}
In the simplified resummed formula, we eliminate the $\mu$ dependence, which means that the resummation is explicitly RG invariant. When including the $\mathcal{O}(\alpha_s^2)$ terms in the boundaries and expanding the resummed formula to $\mathcal{O}(\alpha_s^2)$, we explicitly verified that all the $\mu_h$, $\mu_j$, $\mu_s$, $\nu_j$ and $\nu_s$ dependence is cancelled. This provides another nontrivial check on the RG consistency. Note that the modification from quadrupole contribution now sits in the hard anomalous dimension. With every ingredient at hand, we are ready to discuss the scale setting and perform the numerical Fourier integration.

\section{N${}^3$LL resummation}\label{sec:resummation}

\subsection{Fixed-order expansion}

\begin{figure}[!htbp]
    \centering
    \includegraphics[scale=1.2]{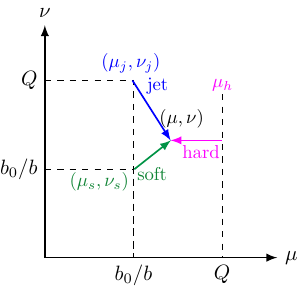}
    \caption{Illustration of all canonical scales and the running.
    The three modes (hard, jet, soft) live at different canonical renormalization and the rapidity scales, while the hard mode does not have a rapidity scale.
    One runs the RG evolutions and rapidty RG evolutions for the three modes to a common arbitrary scale $(\mu,\nu)$. The final result should not depend on $(\mu,\nu)$, as required by the RG consistency.
    }
    \label{fig:scetII_scales}
\end{figure}

\begin{figure}[!htbp]
    \centering
    \includegraphics[scale=0.85]{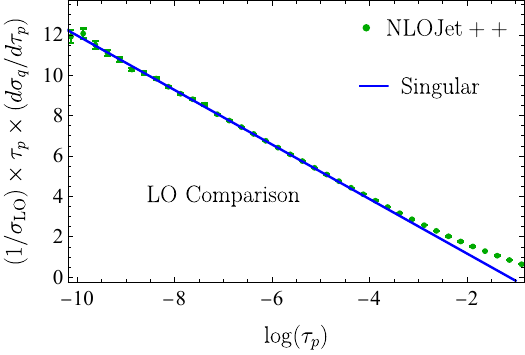}\quad
    \includegraphics[scale=0.89]{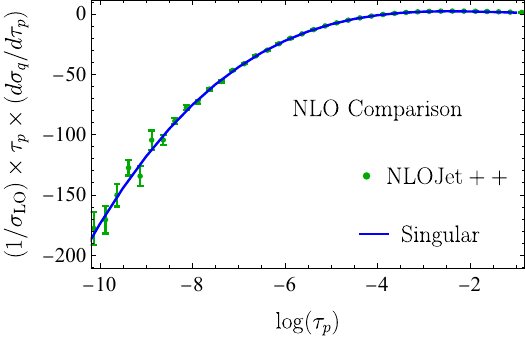}
    \caption{Comparison of the fixed-order expansion of the factorization formula (denoted as ``singular'' in the plots), with the numerical calculations using \texttt{NLOJet++} where two trillion events are sampled.
    We see perfect agreement between the two in the $\tau_p\to 0$ limit, and the error bars show the Monte Carlo errors from \texttt{NLOJet++}.}
    \label{fig:nlojet_compare}
\end{figure}

Before performing the numerical resummation, we need to verify the factorization theorem with the fixed-order data. Note that to eliminate the large logarithms in the hard, jet, and soft functions, we identify the following canonical scales (also illustrated in Fig.~\ref{fig:scetII_scales})
\begin{equation}
    \mu_h=Q,\quad \mu_j=\frac{b_0}{b_y},\quad \mu_s=\frac{b_0}{b_y},\quad \nu_j=Q,\quad \nu_s=\frac{b_0}{b_y}\,.
\end{equation}
Plugging to the resummed expression in Eq.~\eqref{eq:Finalresum}, expanding in terms of $\alpha_s$ and performing the inverse Fourier transformation, we find at $\mathcal{O}(\alpha_s)$
\begin{align}\label{eq:can_naive}
    \frac{1}{\sigma_{\text{LO}}}\frac{d\sigma_q}{d\tau_p}&=\left(\frac{\alpha_s}{4\pi}\right)\int_T \df v \df w \left(\frac{v^2-2 v+w^2-2 w+2}{v w}\right)\bigg\{-16 \left(C_A+2 C_F\right)\frac{ \ln \left(\frac{2 \tau _p}{\xi  Q}\right)}{\tau _p}\nn\\
    &+\left(8 C_A \ln \left(\frac{v w}{1-v-w}\right)-\frac{44 C_A}{3}+16 C_F \ln (1-v-w)-24 C_F+\frac{16 T_F
   n_F}{3}\right)\frac{1}{\tau _p}\bigg\}\nn\\
   &+\mathcal{O}(\alpha_s^2)\,,
\end{align}
where the $v,\, w$ integrals can be performed numerically. To calculate the Fourier integrals analytically, we use the transformation rules between logarithms in Fourier space and plus distributions in the momentum space, summarized in~\app{fouriertransf}.
The order $\alpha_s^2$ expansion is rather long and we provide it in~\app{resum_fo}.
To crosscheck the double and single logarithms, we compare this expansion result with both LO and NLO distributions calculated using \texttt{NLOJet++}~\cite{Nagy:1998bb}. The comparison shows good agreements, see Fig.~\ref{fig:nlojet_compare}.

\subsection{Resummation}

To perform the resummation, we need to freeze the lowest scales to avoid the Landau pole. So instead of Eq.~\eqref{eq:can_naive}, we use the following canonical scale for numerical setup,
\begin{equation}
    \mu_h^{\text{can}}=Q,\quad \mu_j^{\text{can}}=\sqrt{\frac{b_0^2}{b_y^2}+\frac{b_0^2}{b_{\text{max}}^2}},\quad \mu_s^{\text{can}}=\sqrt{\frac{b_0^2}{b_y^2}+\frac{b_0^2}{b_{\text{max}}^2}},\quad \nu_j^{\text{can}}=Q,\quad \nu_s^{\text{can}}=\sqrt{\frac{b_0^2}{b_y^2}+\frac{b_0^2}{b_{\text{max}}^2}}\,,
\end{equation}
where $b_0=2 e^{-\gamma_E}$ and $b_{\text{max}}=1~\text{GeV}^{-1}$. This prescription takes the jet, soft and rapidity soft scales to $\frac{b_0}{b_{\text{max}}}$ when $b_y\rightarrow \infty$. We also do the same treatment for the $\gamma_r\left[\alpha_s(b_0/b_y)\right]$ in the RG kernel in Eq.~\eqref{eq:Wkernel}. 

Then we need to perform three numerical integrals as shown in Eq.~\eqref{eq:Finalresum}. For the $u,\, v$ integrals, we use the Monte Carlo integrator implemented in \texttt{Cuba}~\cite{Hahn:2004fe}. Note that the $y_3$ resolution has cut off the region where $v\to 0$, $w\to 0$ or $u=1-v-w\to 0$, so no subtraction is needed for these two integrals. For the Fourier integral $b_y$, although we resolve the Landau pole, the direct integration is still unstable. There are two approaches to improve it. The first one (approximate method) is to dress the integration with a suppression factor:
\begin{equation}
    \label{eq:approximate_method}
    \Theta_{\text{cut}}=\theta(b_{\text{cut}}-b_y)+\theta(b_y-b_{\text{cut}}) e^{- \eta b_y},\quad \text{with } b_{\text{cut}}=2 \text{ GeV}^{-1},\, \eta=1 \text{ GeV}\,,
\end{equation}
such that the integration can be performed using Monte Carlo methods which exhibit rapid convergence.  There is no canonical choice for the values of $b_{\text{cut}}$ and $\eta$. As we will show below, however, the difference is tiny. The second approach (exact method), which is more rigorous, is to use Fourier integrators, e.g. \texttt{QAWF} method~\cite{KahanerD.K1983Q} or \texttt{DoubleExponential} method~\cite{OOURA1991353}. In this calculation, we adopt the \texttt{QAWF} implemented in \texttt{GSL}~\cite{galassi2018scientific} library for convenience. At N${}^3$LL, since the two-loop hard function contains generalized polylogarithms, we use \texttt{FastGPL}~\cite{Wang:2021imw} for their numerical evaluations.

\begin{table}[!ht]
\begin{center}
\begingroup
\renewcommand{\arraystretch}{1.2}
  \begin{tabular}{|c|c|c|c|c|c|c|}  \hline
   resummation order &     $\Gamma_{\text{cusp}}$ &    $\gamma_{j,s,r}$ &  boundary & $\beta[\alpha_s]$ &  accuracy & non-singular\\  
    \hline
    LL  & 1-loop & -- & tree & 1-loop & $\alpha_s^n L^{2n}$ & -- \\
    \hline
    NLL  & 2-loop & 1-loop & tree & 2-loop & $\alpha_s^n L^{2n-1}$  & -- \\
    \hline
    NNLL & 3-loop & 2-loop & 1-loop & 3-loop & $\alpha_s^n L^{2n-3}$ & LO$_{\text{EEEC}}$ \\
    \hline
    N${}^3$LL & 4-loop & 3-loop & 2-loop & 4-loop & $\alpha_s^n L^{2n-5}$ & NLO$_{\text{EEEC}}$ \\
    \hline
\end{tabular}
\endgroup
\end{center}
\vspace{-0.2cm}
\caption{Definition of the resummation order and their corresponding ingredients~\cite{Becher:2014oda}.}
\label{tab:ords}
\end{table}

Lastly, we need to specify the order counting. As summarized in Tab.~\ref{tab:ords}, our logarithmic counting follows from the standard counting in SCET since we have normalized the distribution by dividing $\sigma_{\text{LO}}$. 
Again, we emphasize that the LO EEEC distribution starts at $\mathcal{O}(\alpha_s^2)$ excluding the boundary term, and its singular expansion is already at NNLL level. Similarly, NLO logarithms are fully predicted only in the N${}^3$LL resummed result. In contrast, for dijet event shapes (e.g. thrust and $C$-parameter), the first matching order is LL/NLL+LO. After resummation, we will switch the normalization from $\sigma_{\text{LO}}$ to the born cross-section with $y_3$ resolution cut for the final result, i.e.
\begin{equation}
    \sigma_1 = \int \df\text{PS}_3 |\mathcal{M}_{\gamma^{*}\to q\bar qg}|^2 \theta(y_3>y_{\text{cut}})\,.
\end{equation}

\begin{figure}[!htbp]
\centering
    \includegraphics[scale=0.72]{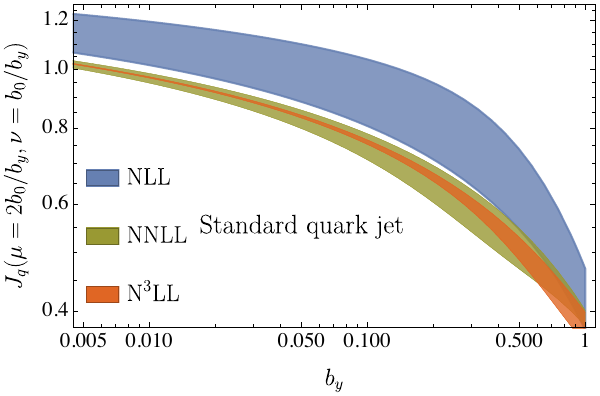}\quad
    \includegraphics[scale=0.72]{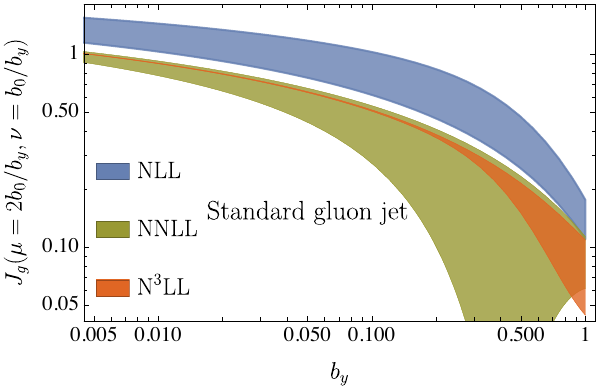}\\
    \includegraphics[scale=0.72]{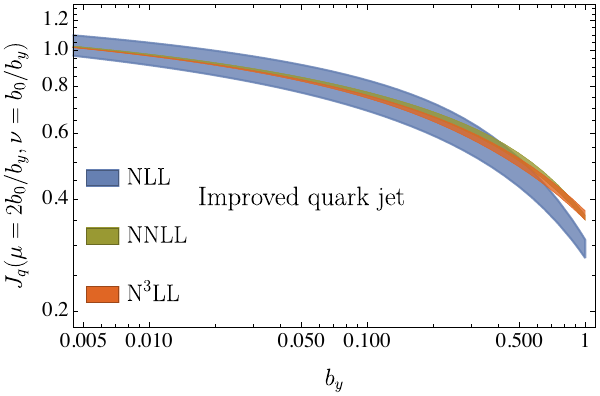}\quad
    \includegraphics[scale=0.72]{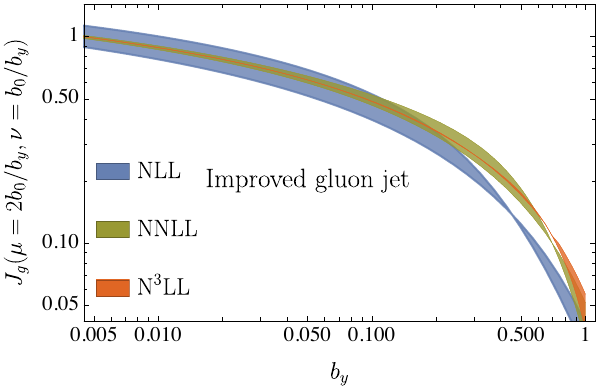}
    \caption{The perturbative convergence of resummed jet function. The upper panel shows the standard jet functions and the lower panel represents our improved jet functions. The uncertainty band is estimated by varying the canonical jet scale $\mu_j^{\text{can}}$ by a factor of 2. Note that $(\mu,\nu)=(2b_0/b_y,b_0/b_y)$ is randomly chosen for illustration; other choices would not change the behavior of bad/good convergence when varying $\mu_j^{\text{can}}$.}
    \label{fig:jetfunction_compare}
\end{figure}

While this order counting convention has been used in many event shape calculations from SCET, we modify the jet function boundaries to improve the perturbative convergence.
In the upper panel of Fig.~\ref{fig:jetfunction_compare}, we use the standard fixed-order expansion of the jet function from Eq.~\eqref{eq:oneloopjet} and evolve the resummed jet function in Eq.~\eqref{eq:jet_resum} from its canonical scale to some randomly chosen scale, e.g., $\mu=2b_0/b_y,\, \nu=b_0/b_y$. The uncertainty bands are obtained by varying the jet scale $\mu_j^{\text{can}}$ by a factor of 2.
We see that the gluon jet function exhibits poor perturbative convergence, as increasing the logarithmic order does not result in a decrease in the uncertainty bands, especially in the relatively large $b_y$ (but still perturbative) regime.
Our investigation shows that the one-loop gluon jet constant $c_g^1$ being numerically too large is responsible for this problem.
To resolve this issue, we prompt $\left(\frac{\alpha_s}{4\pi}\right)c_g^1$ to be an order $\mathcal{O}(\alpha_s^0)$ number and define $e_i^1=1+\left(\frac{\alpha_s(\mu_j)}{4\pi}\right)c_i^1$. Replacing all $c_i^1$ with $\frac{e_i^1-1}{\alpha_s(\mu_j)/(4\pi)}$, we write the improved jet functions to be
\begin{align}\label{eq:jet_imp}
    J_q^{\text{imp.}}&=e_q^1+\left(\frac{\alpha_s}{4\pi}\right)\left[-2C_F e_q^1 L_b L_\omega +L_b\left(3C_F e_q^1+\left(\frac{11}{3}C_A-\frac{4}{3}T_f n_f\right)(e_q^1-1)\right)\right]+\mathcal{O}(\alpha_s^2)\,,\notag\\
    J_g^{\text{imp.}}&=e_g^1+\left(\frac{\alpha_s}{4\pi}\right)\left[-2C_A e_g^1 L_b L_\omega +L_b\left(\left(\frac{11}{3}C_A-\frac{4}{3}T_f n_f\right)(2e_g^1-1)\right)\right]+\mathcal{O}(\alpha_s^2)\,.
\end{align}
We then replace the standard jet function $J_i(\mu_j,\nu_j)$ in \eq{jet_resum} by the improved ones $J_i^{\text{imp.}}(\mu_j,\nu_j)$, and again vary $\mu_j^{\text{can}}$ by a factor of 2.
We see a much better perturbative convergence as shown in the lower panel of Fig.~\ref{fig:jetfunction_compare}.
Therefore, we will use the improved jet functions instead of the standard ones for the final resummation.

\begin{figure}[!htbp]
    \centering
    \includegraphics[scale=0.65]{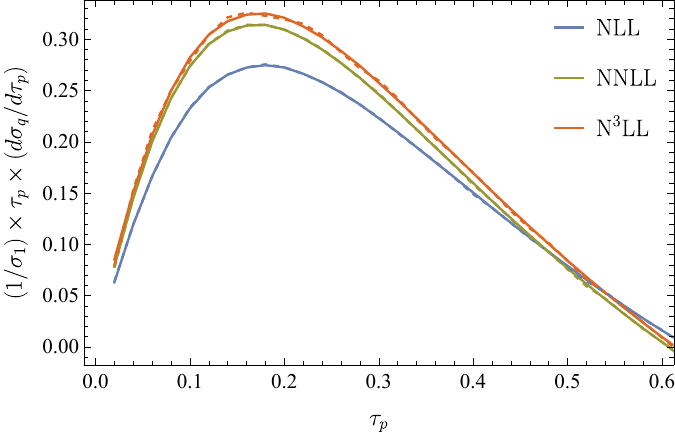}
    \includegraphics[scale=0.67]{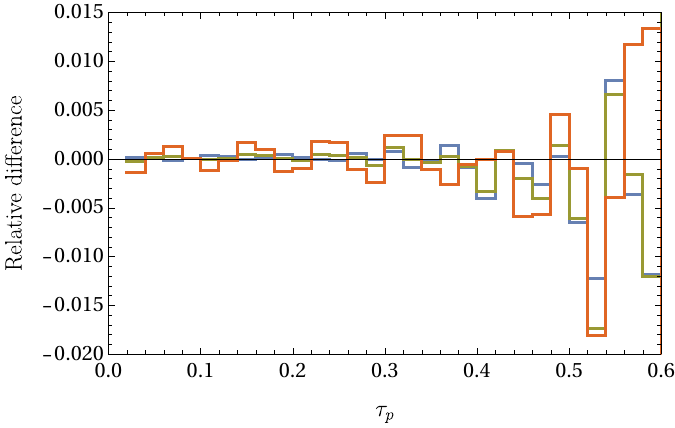}
    \caption{Left: Pure resummation to N${}^3$LL. The solid lines use the exact method and the dashed lines use the approximate method. Both methods give consistent results. Right: the relative difference between the solid and dashed lines on the left. The errors are within $2\%$ as expected.}
    \label{fig:method_compare}
\end{figure}

With the above numerical setup, we calculate the central value of the resummation to N$^3$LL using both approximate and exact methods. Specifically, we set the relative error in \texttt{QAWF} to be $5\times 10^{-2}$ to reduce computational resource requirements. We show these two results in Fig.~\ref{fig:method_compare}, where the solid lines represent the result with the exact method and the dashed lines are the predictions from the approximate method. Both approaches give consistent distributions for all resummation orders. Moreover, we also calculate the relative difference, i.e. $\mathrm{approximate}/\mathrm{exact}-1$, and show that it is within the error setting.

Evaluating the Fourier integral $b_y$ with \texttt{QAWF} requires a large amount of resources since we also have a two-dimensional Monte Carlo sampling for $u$ and $v$. According to the aforementioned comparison, we can safely use the approximate method for the resummations.

\subsection{Scale uncertainties}\label{sec:scale_uncertainty}
In this subsection, we study the scale uncertainties in order to check perturvative convergence and to estimate the higher-order effects. In practice, we introduce five parameters $e_h$, $e_j$, $e_s$, $e_{rj}$ and $e_{rs}$ to vary the hard, jet, soft, rapidity jet and rapidity soft scales, and they all range from $0.5$ to $2$. This amounts to modifying the canonical scales as follows:
\begin{align}
\label{eq:canonical_scales}
    &\mu_h^{\text{can}}=e_h Q,\quad \mu_j^{\text{can}}=e_h e_j\frac{b_0}{b_y} \sqrt{1+\frac{b_y^2}{b_{\text{max}}^2}\frac{1}{e_h^2 e_j^2}}, \quad \mu_s^{\text{can}}=e_h e_s\frac{b_0}{b_y} \sqrt{1+\frac{b_y^2}{b_{\text{max}}^2}\frac{1}{e_h^2 e_s^2}},\nn\\
    &\nu_j^{\text{can}}=\mu_h^{\text{can}} e_{rj},\quad \nu_s^{\text{can}}=e_h e_{rs}\frac{b_0}{b_y} \sqrt{1+\frac{b_y^2}{b_{\text{max}}^2}\frac{1}{e_h^2 e_{rs}^2}}
\end{align}
Note that the variation parameters inside the square root are to make sure $\mu_{j},\mu_s,\nu_{s}\to \frac{b_0}{b_{\text{max}}}$ when $b_y\to\infty$ no matter what the variation is. In other words, since $b_{\text{max}}$ is introduced to avoid the Landau pole and to control the non-perturbative corrections, it should not be affected by the perturbative uncertainty estimation. We also emphasize that the hard variation $e_h$ is correlated among all the scales.

The scale variations follow the Higgs production calculation in Ref.~\cite{Stewart:2013faa}. The first type of variation is the hard scale $e_h$, where we vary all the scales by a factor of 2. The second type of variation is the resummation uncertainty. We choose all combinations of $e_j$, $e_s$, $e_{rj}$ and $e_{rs}$ while keeping $e_h=1$. This generates $3^4-1=80$ possible variations. However, there are certain combinations that double count the logarithm variations. For example, for $\ln\frac{\mu_s}{\nu_j}$ in Eq.~\eqref{eq:Wkernel}, $e_{s}=2, e_{rj}=0.5$ gives a factor of 4. To avoid double counting, we require the variations to have
\begin{equation}
    \left|\log_2\frac{e_s}{e_j}\right|,\, \left|\log_2\frac{e_s}{e_{rs}}\right|,\, \left|\log_2\frac{e_{rj}}{e_{rs}}\right|\leq 1\,.
\end{equation}
In addition, as pointed out in Ref.~\cite{Stewart:2013faa}, 
the rapidity logarithms in Eq.~\eqref{eq:Wkernel} always show up in the combination of
\begin{equation}
    \exp\left[(2C_F+C_A)\ln\frac{\nu_j^2}{\nu_s^2}\left(\frac{1}{2}\gamma_r\left[\alpha_s\left(b_0/b_y\right)\right]-A_\Gamma(\mu_j,b_0/b_y)\right)\right]
    \,.
\end{equation}
Then a simultaneous variation of $e_j=0.5$ with either $e_{rj}=2$ or $e_{rs}=0.5$ has already been accounted by the $\ln\frac{\nu_j}{\nu_s}$ variation. So we exclude these two combinations. Putting together, we get 35 scale variations and the final resummation uncertainty is the envelope of them.

\begin{figure}[!htbp]
    \centering
    \includegraphics[scale=0.68]{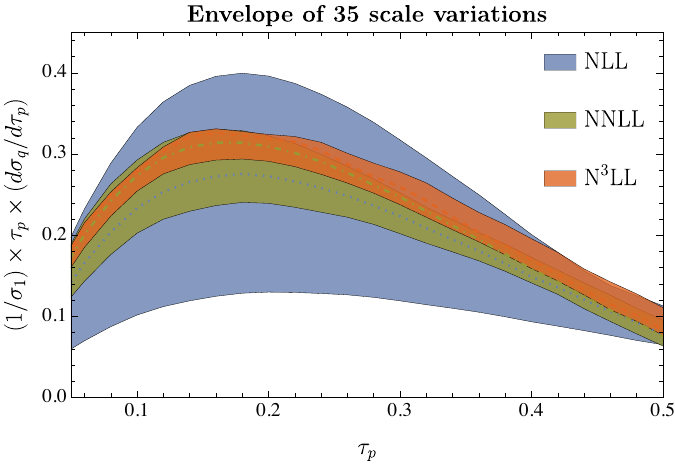}\quad 
    \includegraphics[scale=0.68]{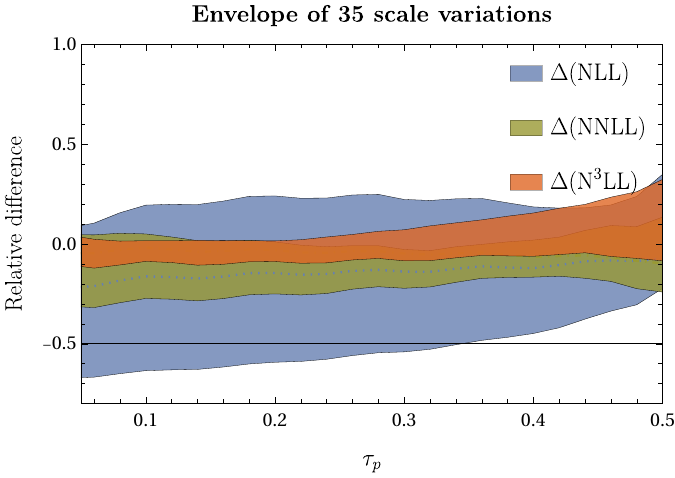}
    \caption{Left panel: final resummation result up to N${}^3$LL accuracy with good perturbative convergence. Right panel: relative difference $\Delta(\text{N}^k\text{LL})\equiv\text{N}^k\text{LL}/\text{N}^3\text{LL}-1$.}
    \label{fig:resum_noMatch_canonical}
\end{figure}

In the left panel of Fig.~\ref{fig:resum_noMatch_canonical}, we show the pure resummation result from NLL to N${}^3$LL for the coplanar EEEC. Increasing the resummation order reduces the uncertainty band as expected and the larger uncertainty from the lower-order resummation covers the higher-order band, indicating good perturbative convergence. In the right panel, we also show their relative difference to the best central curve, i.e. N${}^3$LL prediction to illustrate the convergence.

In principle, we also need to match the perturbative resummation to the fixed-order predictions, i.e. add the non-singular contribution, and perform the renormalon subtraction. Particularly, to avoid double counting, a profile scale~\cite{Abbate:2010xh} is required to smoothly transform the resummation region into the fixed-order region. We will leave these for future work.

\section{Resummation for three gluon jets}
\label{sec:gluon_channel}
In this section, we study the resummation effects for the three gluon jets case, namely contributions initiated from the partonic process $e^+\,e^-\,\to \gamma^*\to\,ggg$. Unlike the $q\bar q g$ case, the three-gluon case is a loop-induced process and starts at two orders higher, i.e., $\mathcal{O}(\alpha_s^2)$ suppressed compared to the quark channel.

The factorization theorem is similar to Eq.~\eqref{eq:full}, with the hard, fragmentation and soft functions replaced with those for the $ggg$ case.
The resummed formula then follows Eq.~\eqref{eq:Finalresum}:
\begin{align}\label{eq:Finalresum_ggg}
    \frac{1}{\sigma_{\text{LO}}} \frac{\df\sigma_g}{\df\tau_p} &= \int_T\! \df v\, \df w\, H_g(v,w,\mu_h)\int_{-\infty}^{\infty}\frac{\df b_y}{2 \pi \xi} \, 2\cos\left(\frac{b_y\tau_p}{\xi}\right)\nn\\
    &\times S_g(\mu_s, \nu_s) J_g (\mu_j, \nu_j) J_g (\mu_j, \nu_j) J_g (\mu_j, \nu_j)\times W_g(\mu_h, \mu_s, \nu_s,\mu_j, \nu_j)\,,
\end{align}
and the RG kernel
\begin{align}\label{eq:Wkernel_ggg}
    W_g(\mu_h, \mu_s, \nu_s,\mu_j, \nu_j) &= \exp\left[6C_A(S_\Gamma(\mu_h,\mu_j)-S_\Gamma(\mu_s,\mu_j))\right]\\
    &\times \exp\left[-2A_{\gamma_{h,g}}(\mu_h,\mu_s)+2A_{\gamma_{J,g}}(\mu_s,\mu_j)+2A_{\gamma_{J,g}}(\mu_s,\mu_j)+2A_{\gamma_{J,g}}(\mu_s,\mu_j)\right]\nn\\
    &\times\left(\frac{Q^2}{\mu_h^2}\right)^{3C_A A_\Gamma(\mu_j,\mu_h)}\times \left(\frac{\nu_j^2}{\nu_s^2}\right)^{\frac{3C_A}{2}\gamma_r\left[\alpha_s\left(b_0/b_y\right)\right]-3C_A A_\Gamma(\mu_j,b_0/b_y)}\nn\\
    &\times \left(\frac{\mu_s^2}{\nu_s^2}\right)^{3 C_A A_\Gamma(\mu_j,\mu_s)}\times \left[(1-w)(1-v)\right]^{2C_A A_\Gamma(\mu_j,\mu_s)}(v+w)^{2C_A A_\Gamma(\mu_j,\mu_s)}\,.\nn
\end{align}

To have a consistent logarithmic counting in both channels, we start by building up some conventions below. First of all, if we define
\begin{equation}
    r_q(\tau_p)\equiv \frac{1}{\sigma_{\text{LO}}}\frac{d\sigma_q}{d\tau_p},\quad r_g(\tau_p)\equiv \frac{1}{\sigma_{\text{LO}}}\frac{1}{\alpha_s^2}\frac{d\sigma_g}{d\tau_p}
\end{equation}
then the distribution strictly follows the counting in Tab.~\ref{tab:ords}, where the N${}^k$LL accuracy in $r_i(\tau_p)$ corresponds to $\alpha^n L^{2n}\sim \alpha_s^n L^{2n-(2k-1)}$. More importantly, we can use the same convention for all ingredients in the $ggg$ channel resummation. Then to obtain the sum of the distribution, $\frac{1}{\sigma_{\text{LO}}}\frac{d\sigma}{d\tau_p}=r_q(\tau_p)+\alpha_s^2 r_g(\tau_p)$, we need to determine what logarithmic order in $r_g(\tau_p)$ is as important as the N${}^k$LL order of $r_q(\tau_p)$. It turns out that up to NNLL accuracy, there is no $ggg$ channel required, and for N${}^{k}$LL accuracy with $k\geq 3$, we need to add N${}^{k-2}$LL resummation from $ggg$ channel. That is to say, our N${}^3$LL resummation in the $q\bar qg$ case requires corrections from NLL in $ggg$ channel.

\begin{figure}[!htbp]
    \centering
    \includegraphics[scale=0.4]{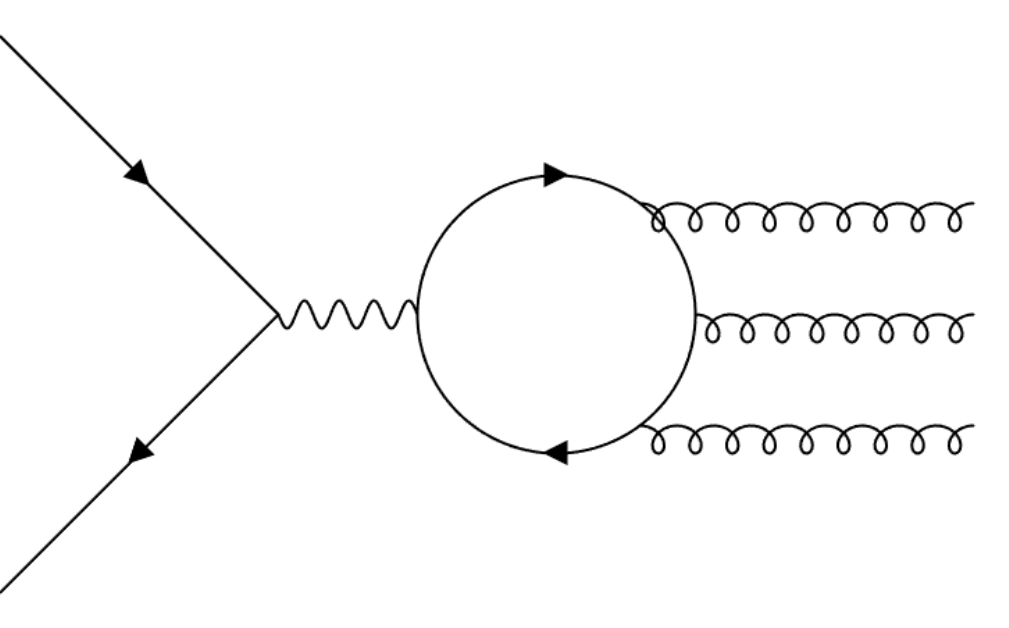}
    \caption{An example diagram of the $e^+e^-\to \gamma^* \to ggg$ process. Squaring this amplitude results in an $\cO(\alpha_s^2)$ suppression.
    }
    \label{fig:gggFeynmanDiagrams}
\end{figure}

At NLL, we need the two-loop cusp anomalous dimension, one-loop regular hard, jet, and soft anomalous dimensions, and the LO three-gluon hard function. First of all,
the LO hard function is simply the first order differential cross section for $e^+e^-\to ggg$~\cite{Laursen:1981zu,Baier:1980}.
An example Feynman diagram can be found in Fig.~\ref{fig:gggFeynmanDiagrams}. The hard function for $e^+ e^- \to ggg$ satisfies the same RG equation as shown in Eq.~\eqref{eq:hard_RGE}. By evaluating the right-hand side of Eq.~\eqref{eq:hard_RGE} for three gluon final states, i.e., 
\begin{align}
\label{eq:color3g}
    &\mathbf{T}_{1} \cdot \mathbf{T}_2 =  \mathbf{T}_{1} \cdot \mathbf{T}_3 =  \mathbf{T}_{2} \cdot \mathbf{T}_3 = -\frac{1}{2}\mathbf{T}_1^2 = -\frac{1}{2} \mathbf{T}_2^2 = -\frac{1}{2} \mathbf{T}_3^2 = -\frac{1}{2}C_A, \nonumber \\
    &  f_{abe}f_{cde} \, \sum_{i=1}^3\sum_{\substack{{1\leq j<k\leq 3}\\ j,k\neq i}}\bigl\{\bT_i^a,  \bT_i^d\bigr\}   \bT_j^b \bT_k^c
    \,\bigl\lvert gg g\bigr\rangle\,=\,  9 C_A \bigl\lvert g g g\bigr\rangle\,
    \,,
\end{align}
the Eq.~\eqref{eq:hardRGE_cuspNcusp} becomes
\begin{equation}
\label{eq:hardRGE_cuspNcuspGGG}
    \frac{\df\ln H_{g}(v,w,\mu)}{\df\ln\mu^2} = \frac{3 C_A}{2} \gamma_{\rm cusp}(\alpha_s) \ln \frac{Q^2}{\mu^2}  + \gamma_{h,g}(v,w,\alpha_s) \,,
\end{equation}
where the non-cusp piece for three gluons is
\begin{align}\nn
  \gamma_{h,g}(v,w,\alpha_s) =\, &\gamma_{\rm cusp}(\alpha_s)\frac{C_A}{2} \left( \ln v + \ln w + \ln(1-v-w) \right)\nn\\
  &+3 \gamma_g(\alpha_s)
  \, +9 \,C_A\gamma_{\rm quad}[\alpha_s]
 \,.
\end{align}
We emphasize that the $\gamma_{\text{quad}}(\alpha_s)$ starts at $\mathcal{O}(\alpha_s^3)$ and will not contribute until N${}^3$LL.

\begin{figure}[!htbp]
    \centering
    \includegraphics[scale=0.7]{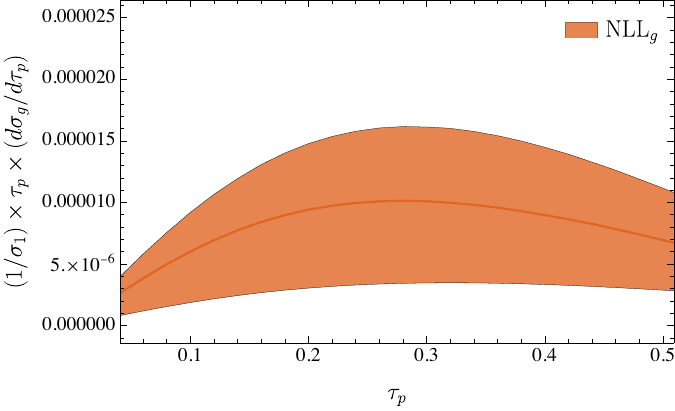}
    \includegraphics[scale=0.7]{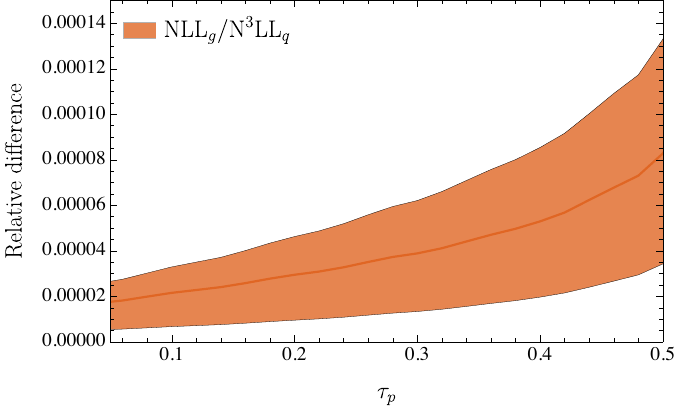}
    \caption{Left: NLL resummation for $e^+e^-\to \gamma^*\to ggg$ process. Right: the relative difference between this NLL resummation and the N${}^3$LL resummation of the $q\bar q g$ channel. Note that because the $q\bar q g$ cross-section is close to zero near $\tau_p=0.5$, the relative difference starts to increase.}
    \label{fig:gggresum_noMatch_canonical}
\end{figure}

The soft function for the three gluons process has the same form as in Eq.~\eqref{eq:softFunctionForm}, and satisfies the same RG equation in Eq.~\eqref{eq:RG_soft_UV}-\eqref{eq:RG_soft_rap}. Using Eq.~\eqref{eq:color3g} and Eq.~\eqref{eq:nij}, the soft RG in Eq.~\eqref{eq:softRGE}-\eqref{eq:rapidityRGE} become 
\begin{align}
  \label{eq:softRGEggg}
  \frac{\df\ln S_{g}}{\df\ln \mu^2} =\, & \frac{3 C_A }{2} \left(\gamma_{\rm cusp}[\alpha_s] \ln \frac{\mu^2}{\nu^2} - \gamma_s[\alpha_s] \right) 
+ \frac{C_A}{2} \gamma_{\rm cusp}[\alpha_s] \ln \frac{(v+w)^2 (1-v)^2(1-w)^2 }{vw (1 - v - w)}
-9\,C_A\gamma_{\rm quad}[\alpha_s]
 \,,\\
  \label{eq:rapidityRGEggg}
  \frac{\df\ln S_{g}}{\df\ln \nu^2} =\,& \frac{3 C_A}{2}  \left( \int_{\mu^2}^{b_0^2/b_y^2} \frac{\df\bar \mu^2}{\bar \mu^2} \gamma_{\rm cusp}(\alpha_s[\bar \mu]) + \gamma_r (\alpha_s[b_0/b_y])  \right)  \,.
\end{align}
With all the ingredients at hand, we now perform the Fourier integral in Eq.~\eqref{eq:Finalresum_ggg} with the same setup in the $q\bar q g$ channel, namely the approximate method in Eq.~\eqref{eq:approximate_method}. We also use the envelope of 35 variations to estimate the uncertainty band. The distribution is normalized to $\sigma_1$ for consistency as in the $q\bar q g$ channel. In Fig.~\ref{fig:gggresum_noMatch_canonical}, we show both the NLL resummation including the perturbative uncertainty band and its ratio to the N${}^3$LL resummation in the $q\bar q g$ channel. The contribution from the $ggg$ channel is relatively small: only 0.02\% correction to our highest-order resummation in the range $0<\tau_p\leq 0.5$. This is expected because the hard function is already suppressed by $\alpha_s^2$.
While the comparison suggests we can ignore the $ggg$ channel in the current study, we want to emphasize the existence of this additional channel in N${}^3$LL and beyond, which could become phenomenologically relevant in future scenarios, e.g. when looking at the next-to-leading power corrections. To our knowledge, our result is the first resummation for three gluon jets at $e^+e^-$.

\section{Fully differential EEEC in the coplanar limit}\label{sec:diffspectrum}
In this section, we study the three-particle coplanar limit in the fully differential EEEC as defined in \eq{EC_def} with $n=3$. Below we will briefly review the variables $\{s,\phi_1,\phi_2\}$ and consider the $s\to1$ limit.
Then we will explore the resummation effects of different coplanar configurations, which are characterized by the angles $\{\phi_1,\phi_2\}$.

\begin{figure}[!hbtp]
    \centering
    \includegraphics[scale=0.25]{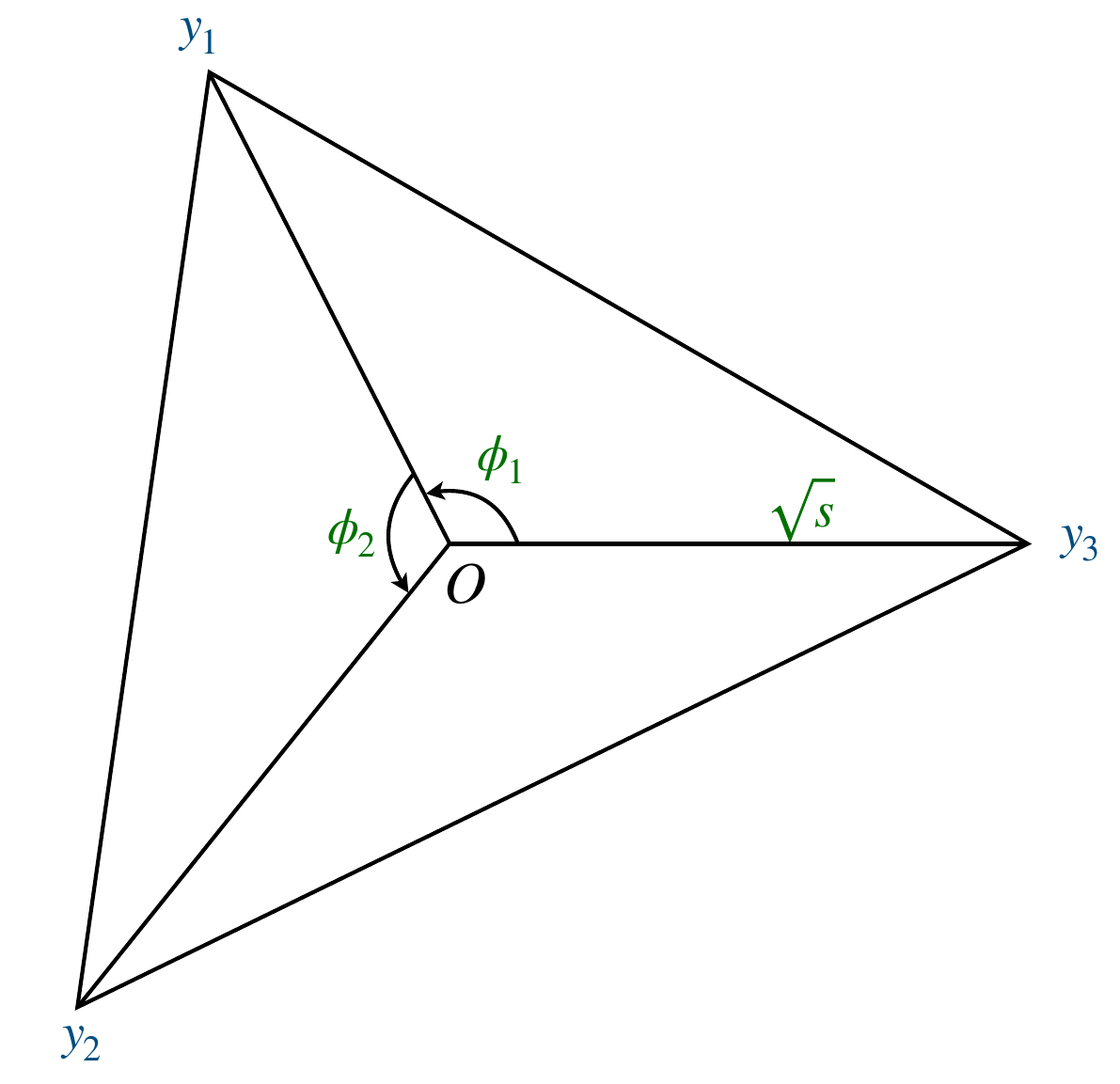}
    \caption{Demonstration of celestial coordinate. By mapping the angular distances to three points $y_{1,2,3}$ on the celestial sphere, we introduce the radius $s$ and two angle variables $\phi_{1,2}$.}
    \label{fig:ypara}
\end{figure}

To begin with, we review the celestial coordinate introduced in Ref.~\cite{Yan:2022cye}. For any three final-state particles, if we label the angles between any pair of them to be $\theta_{ij}$, then we can map them to three points on the celestial sphere:
\begin{equation}
    \frac{1-\cos\theta_{ij}}{2}=\frac{|y_i-y_j|^2}{(1+|y_i|^2)(1+|y_j|^2)},\quad i\neq j\in\{1,2,3\}\,.
    \label{eq:xtoy}
\end{equation}
Using these three points $y_{1,2,3}$, we can construct a circle and define the radius $\sqrt{s}$ as well as two angles $\phi_{1,2}$ for $y_1$ and $y_2$, as shown in Fig.~\ref{fig:ypara}. This amounts to the following formulae
\begin{equation}
    y_1=\sqrt{s} e^{i\phi_1},\quad y_2=\sqrt{s} e^{i(\phi_1+\phi_2)},\quad y_3=\sqrt{s}
    \label{eq:ytotau} \,. 
\end{equation}
On the one hand, $s\rightarrow 0$ will force $\theta_{ij}$ goes to zero homogeneously and approaches the collinear limit.
On the other hand, in the $s\to 1$ limit, the three particles fall onto the same plane, with $\theta_{13}=\phi_1$, $\theta_{21}=\phi_2$, $\theta_{32}=2\pi-\phi_1-\phi_2$.
Large logarithms of $\ln(1-s)$ in the EEEC were noticed through fixed-order calculations~\cite{Yan:2022cye,Yang:2022tgm,Yang:2024gcn}.
In particular, the leading-power contribution in the three-particle coplanar limit ($s\to1$) with fixed $\phi_1$, $\phi_2$ is governed by the trijet coplanar configuration where $\phi_1$, $\phi_2$ characterize the relative angles between pairs of the three jets\footnote{Notice that the trijet configuration requires $\phi_1$, $\phi_2$ satisfy $\phi_1<\pi$, $\phi_2<\pi$, $\phi_1+\phi_2>\pi$. Therefore, there is no large $\ln(1-s)$ in regions outside $\phi_1<\pi$, $\phi_2<\pi$, $\phi_1+\phi_2>\pi$, as noticed in Ref.~\cite{Yang:2024gcn}.}.
Therefore, in this limit, $\{\tau,u,v\}$ and $\{s,\phi_1,\phi_2\}$ are interchangeable, and we can convert the factorization theorem in the $\tau_p\to0$ limit to the one for $s\to1$ limit.
We find
\begin{align}\label{eq:coordinate_convert}
    \tau_p&=2(1-s)\sin\frac{\phi_1}{2}\sin\frac{\phi_2}{2}\sin\frac{(\phi_1+\phi_2)}{2}\,,\nn\\
    \{u,v,w\}\in\{z_i\}
    &=\Bigg\{\cot\frac{\phi_1}{2}\cot\frac{\phi_2}{2},-\cot\frac{\phi_1}{2}\cot\frac{(\phi_1+\phi_2)}{2},-\cot\frac{\phi_2}{2}\cot\frac{(\phi_1+\phi_2)}{2}\Bigg\}\,,
\end{align}
where $u,v,w$ can take any order of the three terms $\{z_i\}$ from the right-hand side, corresponding to the $S_3$ symmetry of EEEC. From \eq{coordinate_convert}, we can write $\xi$ in \eq{xi} as
\begin{equation}
    \xi = \frac{4}{Q}\sin\frac{\phi_1}{2}\sin\frac{\phi_2}{2}\sin\frac{(\phi_1+\phi_2)}{2}=\frac{2\tau_p}{Q(1-s)}
    \,.
\end{equation}

Therefore, in order to derive the factorization theorem for the fully differential EEEC in the $s\to1$ limit, 
we can convert from the factorization formula for $\tau_p$ in \eq{factexp} rather than starting from scratch.
First, we account for the Jacobian factor
\begin{align}
    \left|\frac{\partial(\tau_p,u,v)}{\partial(s,\phi_1,\phi_2)}\right|=-\cot\frac{\phi_1}{2}\cot\frac{\phi_2}{2}\cot\frac{(\phi_1+\phi_2)}{2}
    \,.
\end{align}
In addition, instead of normalizing by the jet energies $E_{J_1}E_{J_2}E_{J_3}$ as in \eq{EEEC_cpln_def}, here the normalization factor is $Q^3$, which results in an extra factor
\begin{align}\nn
    \frac{E_{J_1}E_{J_2}E_{J_3}}{Q^3}&=-\frac{\sin\phi_1\,\sin\phi_2\,\sin(\phi_1+\phi_2)}{\bigl(\sin\phi_1+\sin\phi_2-\sin(\phi_1+\phi_2)\bigr)^3}
    \\[.15cm]
    &=-\frac{1}{8}\cot\frac{\phi_1}{2} \cot\frac{\phi_2}{2} \cot\frac{(\phi_1+\phi_2)}{2} \csc\frac{\phi_1}{2} \csc\frac{\phi_2}{2} \csc\frac{(\phi_1+\phi_2)}{2}\,.
\end{align}
Accounting for all these factors and the $S_3$ symmetry, we obtain the factorization formula for the $q\bar q g$ channel in the $s\to 1$ limit for the fully differential spectrum,
\begin{align}
    \frac{1}{\sigma_{\text{LO}}}\frac{d^3\sigma_q}{\df s \df \phi_1 \df\phi_2} &=\frac{Q}{8}\Bigg[\cot\frac{\phi_1}{2} \cot\frac{\phi_2}{2} \cot\frac{(\phi_1+\phi_2)}{2} \csc\frac{\phi_1}{2} \csc\frac{\phi_2}{2} \csc\frac{(\phi_1+\phi_2)}{2}\Bigg]^2\times\sum_{i,j,k} H_q(z_i,z_j,z_k,\mu)\nn\\
    &\times \int_0^\infty \frac{\df b_y}{2\pi} \cos\left(\frac{b_y Q (1-s)}{2}\right)\times S_q(b_y,\mu,\nu)J_{q}(b_y,\mu,\nu)J_{\bar q}(b_y,\mu,\nu)J_{g}(b_y,\mu,\nu)
    \,,
\end{align}
where $i,j,k$ run over all six permutations and we emphasize that there is $z_i$ dependence in the hard anomalous dimensions. The $ggg$ channel again takes a similar factorized form with associated ingredients. Solving RGEs for hard, jet, and soft functions are identical to the $\tau_p$ case and so is the final resummed form, therefore we don't write them down here. 

The advantage of studying a fully differential spectrum is that setting different values of $\phi_{1,2}$ allows us to study different coplanar configurations (or trijet shapes). For example, $\phi_1=\phi_2=2\pi/3$ represents the symmetric trijet configuration and $\phi_1\to 0$ or $\phi_2\to 0$ or $\phi_1+\phi_2\to \pi$ corresponds to the squeezed limit, where there are two jets collinear with each other and the third jet being backward.

\begin{figure}[!htbp]
    \centering
    \includegraphics[scale=0.65]{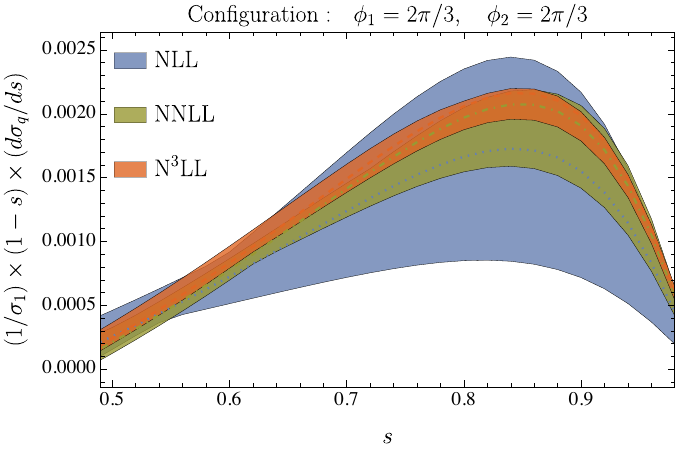}\quad
    \includegraphics[scale=0.65]{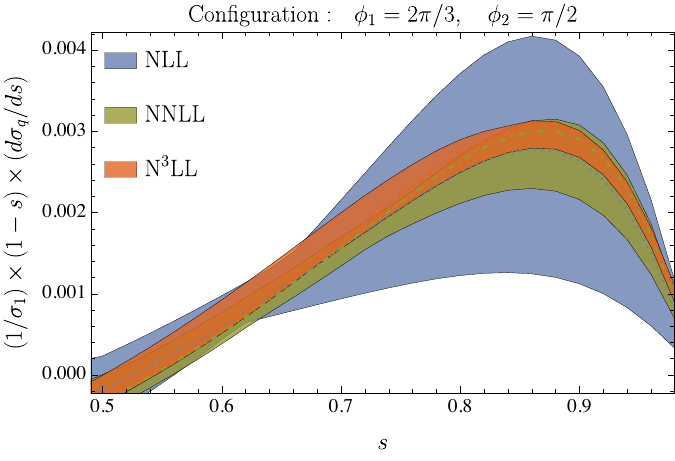}\\
    \includegraphics[scale=0.65]{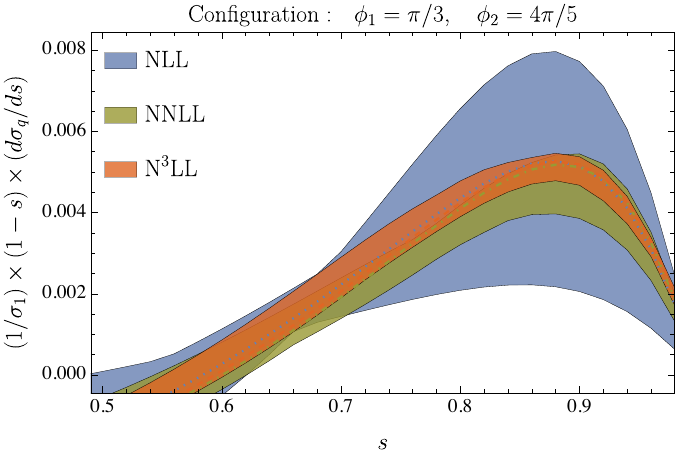}\quad
    \includegraphics[scale=0.65]{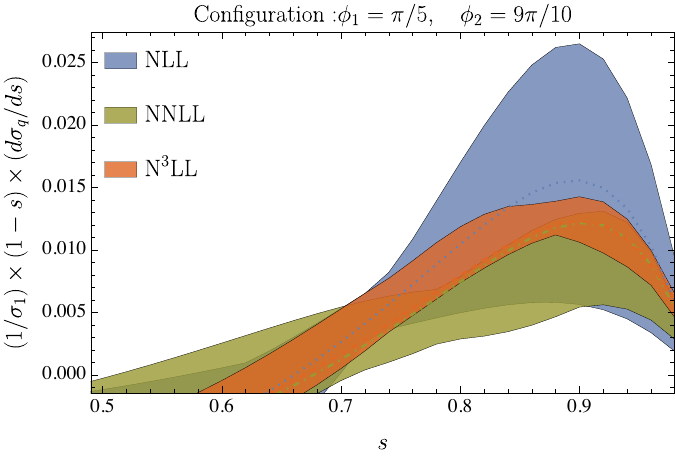}
    \caption{Resummation of fully differential spectrum with four configurations. For each configuration, we find good convergence in the resummation orders.}
    \label{fig:resum_s_configuration}
\end{figure}

In Fig.~\ref{fig:resum_s_configuration}, we show the resummation results for four choices of $\{\phi_1,\phi_2\}$: $\{2\pi/3,2\pi/3\}$, $\{2\pi/3, \pi/2\}$, $\{\pi/3, 4\pi/5\}$ and $\{\pi/5, 9\pi/10\}$. The first one corresponds to the exact symmetric configuration, the second and the third are regular configurations, and the last one approaches the squeezed limit. While the higher-order band mostly falls into the lower-order one in the first three cases, we observed slightly poor convergence in the last configuration. In particular, the distribution goes to negative faster than others, indicating that we should turn off the resummation below $s\sim 0.7$. 
This is reasonable because the cross-section also receives additional logarithms in the squeezed limits, e.g. $\ln \left(\tan^2\frac{\phi_1}{2}\tan^2\frac{\phi_2}{2}\tan^2\frac{\phi_1+\phi_2}{2}\right)$. In principle, these logarithms need to be resummed in the fully differential spectrum but we leave it for future work. On the contrary, in the $\tau_p$ spectrum, the jet algorithm protects us from the squeezed singularities and thus we do not observe these effects. 

\section{Analysis}\label{sec:analysis}
In this section, we perform some further studies.
In \sec{pythia}, we explore the non-perturbative power corrections using the Monte Carlo program $\texttt{Pythia8}$~\cite{Sjostrand:2014zea}.
In \sec{Dparameter}, we show a simple relation between the $\tau_p$-projected EEEC with the $D$-parameter.
\subsection{Non-perturbative corrections}\label{sec:pythia}

So far we are only studying the perturbative predictions for coplanar EEEC, while experiments will measure the observable on hadrons. There are different ways to turn the partonic predictions into hadronic results that can be compared with the experimental data. In general, the non-perturbative power corrections to the event shape observables are suppressed by $\Lambda_{\text{QCD}}/Q$, where $Q$ is the center-of-mass energy. In the current collider programs, many analyses use the Monte Carlo generators to estimate the non-perturbative effects from the hadronization process. On the other hand, there are also analytic approaches, such as renormalon-based calculation~\cite{Beneke:1998ui} and SCET-based OPE analysis~\cite{Lee:2006nr,Hoang:2007vb}, to extract the leading power corrections to a specific QCD observable.

It will be interesting to rigorously study these effects for coplanar EEEC using the operator language in SCET, but in this paper, we first present a qualitative analysis of the hadronization effect using \texttt{Pythia8}~\cite{Sjostrand:2014zea} with \texttt{FastJet} interface~\cite{Cacciari:2011ma}.
In the numerical setup, we sample 10 million events in \texttt{Pythia8}, select the three-jet events with $k_T$ algorithm implemented in \texttt{FastJet}, and calculate coplanar EEEC $\tau_p$ before and after the hadronization. To be consistent, we also set $\alpha_s=0.118$ and the $y_3$ cutoff to be $y_{\text{cut}}=0.1$. To illustrate the size of hadronization correction, we then calculate the ratio difference of the spectrum, i.e. $\Delta(d\sigma/d\tau_p)=(d\sigma^\text{hadron}/d\tau_p)/(d\sigma^\text{parton}/d\tau_p)-1$ bin by bin.

\begin{figure}[!htbp]
    \centering
    \includegraphics[width=0.48\linewidth]{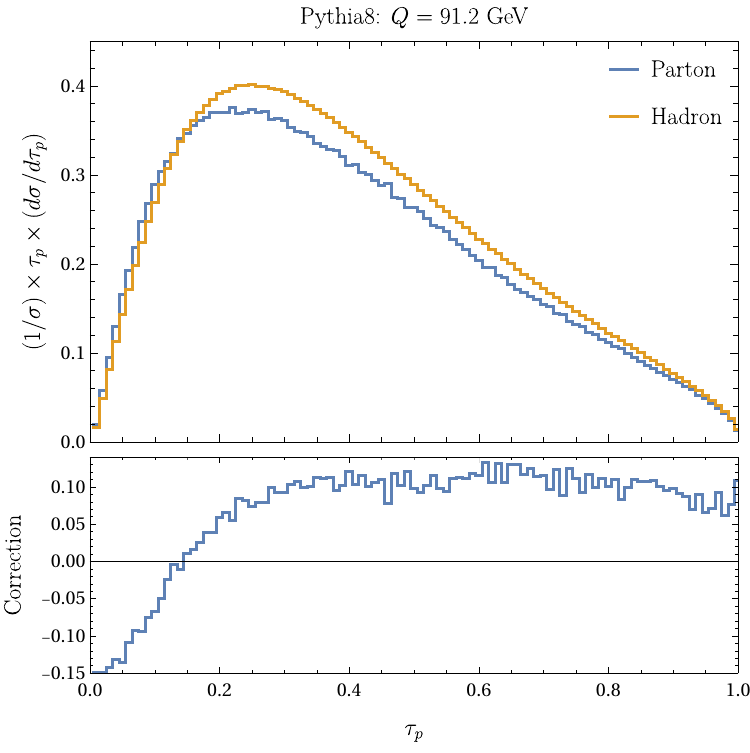}\,
    \includegraphics[width=0.48\linewidth]{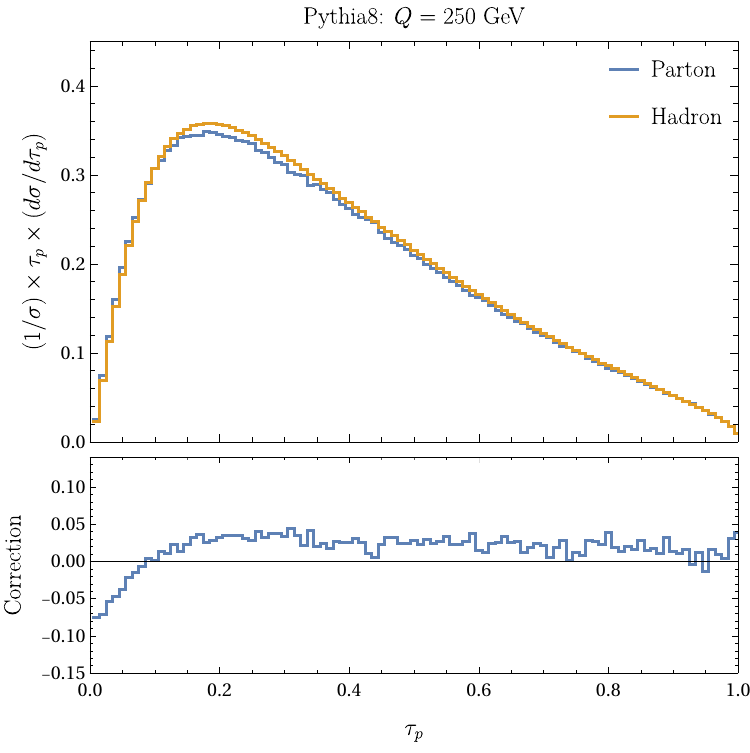}
    \caption{Partonic and hadronic distributions of the coplanar EEEC in \texttt{Pythia8}, with $Q=91.2$ GeV and $Q=250$ GeV.}
    \label{fig:hadronization}
\end{figure}

In Fig.~\ref{fig:hadronization}, we show the distribution from \texttt{Pythia8} at $Q=91.2$GeV and $Q=250$GeV, where the latter corresponds to the energy for CEPC. Because the parton shower and hadronization model include more things than our fixed-order calculation or the analytic resummation, we normalize the \texttt{Pythia8} spectrum by integration of the distribution itself, i.e. instead of $\sigma_1$, we use $\sigma = \int_0^1 d\tau_p( d\sigma^{\text{parton}/\text{hadron}}/d\tau_p)$.
We find that the hadronization correction $\Delta(d\sigma/d\tau_p)$ is less than $15\%$ for $Q=91.2$ GeV and less than $8\%$ for $Q=250$ GeV. In the small $\tau_p$ region, we learn from the coplanar resummation that the smallest scale is $\mu_j=\mu_s=b_0/b_y\sim \tau_p Q$. Then we would qualitatively expect the non-perturbative correction in powers of $\Lambda_{\text{QCD}}/(\tau_p Q)$, which is consistent with the $\tau_p<0.2$ region in the lower panels of Fig.~\ref{fig:hadronization}.

\begin{figure}
    \centering
    \includegraphics[width=0.32\linewidth]{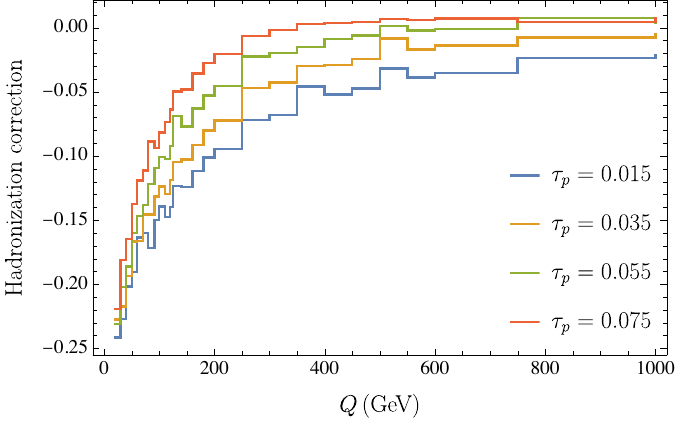}
    \includegraphics[width=0.32\linewidth]{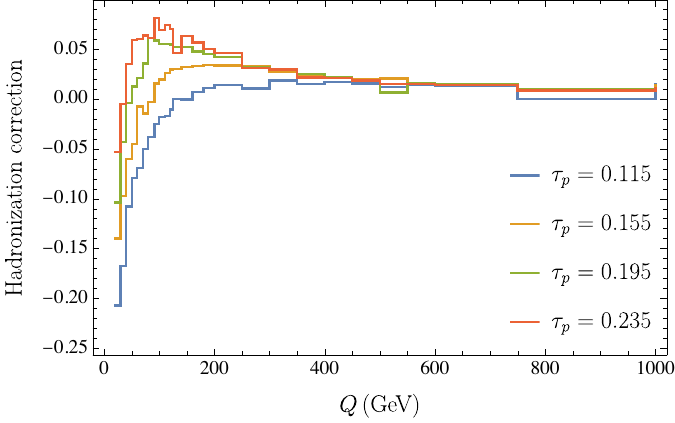}
    \includegraphics[width=0.32\linewidth]{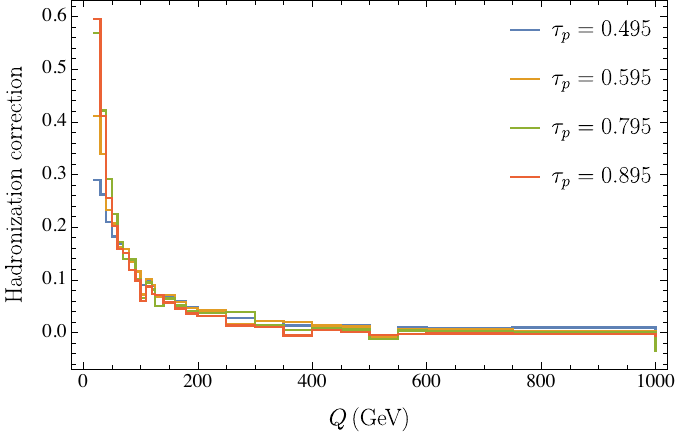}
    \caption{$Q$ dependence of hadronization correction with $\tau_p$ fixed. In these three plots, we focus on the small, middle, and large $\tau_p$ regions of the distribution, and observe a transition of $Q$ dependence in the middle region.}
    \label{fig:hadronization_tau123}
\end{figure}

To study the $Q$ dependence of the hadronization effects, we also choose a few $\tau_p$ values and calculate the hadronization correction $\Delta(d\sigma/d\tau_p)$ with different energies $Q$, ranging from 20 GeV to 1 TeV. As shown in Fig.~\ref{fig:hadronization_tau123}, when $\tau_p$ is small ($\tau_p=0.015,0.035,0.055,0.075$) or large ($\tau_p=0.495,0.595,0.795,0.895$), the size of hadronization correction decreases as the energy $Q$ increases, though they have opposite signs. This is also consistent with the argument that non-perturbative correction is suppressed by powers of $\Lambda_{\text{QCD}}/Q$. However, there exists a transition region around $\tau_p=0.2$ that exhibits different $Q$ dependence. It will be interesting to derive the trijet operator using SCET and study the hadronization effect quantitatively in the Fourier space. We leave them for future work.

\subsection{Relation between $D$-Parameter and $\tau_p$-projected EEEC}\label{sec:Dparameter}
As mentioned in the introduction, the $D$-parameter is also a canonical event shape to study the coplanar trijet configuration.
As an aside, in this subsection, we derive a simple relation between the third moment of the $\tau_p$-projected EEEC defined in \eq{EEEC_tau_def} and the $D$-parameter.
  
In massless QCD, the $C$-parameter and the $D$-parameter are defined as~\cite{Parisi:1978eg, Donoghue:1979vi, Ellis:1980wv},
\begin{equation}
C=3(\lambda_1\lambda_2+\lambda_1\lambda_3+\lambda_2\lambda_3)
\,,\qquad
      D=27\lambda_1\lambda_2\lambda_3\,,
\end{equation}
where $\lambda_i$ are the three eigenvalues of the spherocity tensor $\Theta_{\alpha\beta}$,
\begin{align}
    \Theta_{\alpha\beta} = \frac{1}{Q}\sum_i\frac{k_{i\alpha}k_{i\beta}}{E_i}\,,
\end{align}
Here $i$ runs over all final-state particles. $Q$ here is the center-of-mass energy,   $E_i$ is the energy of particle $i$, and $k_{i\alpha}$ is the $\alpha$ component of particle $i$'s three momentum ($\alpha, \beta=1,2,3$).

Writing the $D$-parameter in terms of $\Tr\,\Theta \,(=1)$, $\Tr\,\Theta^2$ and $\Tr\,\Theta^3$, we find
\begin{align}
        D&=\frac{27}{6}\left\{\left(\lambda_1+\lambda_2+\lambda_3\right)\left[\left(\lambda_1+\lambda_2+\lambda_3\right)^2-3\left(\lambda_1^2+\lambda_2^2+\lambda_3^2\right)\right]+2\left(\lambda_1^3+\lambda_2^3+\lambda_3^3\right)\right\}\nonumber\\
        &=\frac{27}{6}\left\{\Tr\, \Theta\left[\left(\Tr\, \Theta\right)^2-3\,\Tr\, \Theta^2\right]+2\,\Tr\, \Theta^3\right\}\nonumber\\
        &=\frac{27}{6}\left(1-3\sum_{i,j}\frac{E_iE_j\cos^2\theta_{ij}}{Q^2}+2\sum_{i,j,k}\frac{E_iE_jE_k\cos\theta_{ij}\cos\theta_{jk}\cos\theta_{ki}}{Q^3}\right)\nonumber\\
        &=\frac{27}{Q^{3}}\sum_{i<j<k}E_{i} E_{j} E_{k}\left(1+2 \cos \theta_{i j} \cos \theta_{j k} \cos \theta_{i k}-\cos ^{2} \theta_{i j}-\cos ^{2} \theta_{j k}-\cos ^{2} \theta_{i k}\right)\nonumber\\
       &=\frac{27}{Q^{3}} \sum_{i<j<k} \frac{\left|\left(\vec k_i \times \vec k_j\right)\cdot\vec k_k \right|^{2}}{E_{i} E_{j} E_{k}}\nonumber\\
       &=\frac{27}{Q^{3}} \sum_{i<j<k}E_{i} E_{j} E_{k}\tau^2_{ijk}\,.
\end{align}
Therefore, the average of the $D$-parameter is proportional to the second moment of our $\tau_p$-projected EEEC,
    \begin{equation}
        \left<D\right>=\frac{9}{2}\int_0^1\! \df\tau_p ~ \tau_p^2 \frac{1}{\sigma_{\text{tot}}}\frac{\df\sigma}{\df\tau_p}\,,
    \end{equation}
where $\sigma_{\text{tot}}$ is the total cross section.

A similar relation also exists between the $C$-parameter and the EEC (noticed in e.g. Ref.~\cite{Dasgupta:2003iq}). We also derive it here,
\begin{align}
 C &= \frac{3}{2}\left[(\lambda_1+\lambda_2+\lambda_3)^2-(\lambda_1^2+\lambda_2^2+\lambda_3^2)\right]\\
\nonumber   &=\frac{3}{2}\left[ 1-\Tr(\Theta^2) \right]\\
&=\frac32 \biggl(1-\sum_{i,j}\frac{E_iE_j\cos^2\theta_{ij}}{Q^2}\biggr)
\,,
\end{align}
Therefore, we find that the average of the $C$-parameter is related to the moments of the EEEC by
\begin{align}
    \left<C\right>=\frac32\left(1-\int_0^1\!\df x\,(1-2x)^2\, \frac{1}{\sigma_{\text{tot}}}\frac{\df\sigma}{\df x}\right)=6\int_0^1\!\df x\,x(1-x) \,
    \frac{1}{\sigma_{\text{tot}}}\frac{\df\sigma}{\df x}\,,
\end{align}
where $x$ is $x_{12}$ in \eq{EC_def}.

Inspired by these relations, it will be interesting to generalize the factorization theorem developed for the EEC and the coplanar EEEC to the trijet region of both $C$-parameter and $D$-parameter in the future.

\section{Conclusion}\label{sec:conclusion}

Precise measurements for the Standard Model, in particular QCD and the Higgs sector, have been one of the primary tasks in collider physics, for both improving our understanding of quantum field theories and helping with new physics searches. In recent years, energy correlators have opened up a new precision era for both $e^+e^-$ and proton-proton colliders. First of all, the energy correlator has a very simple definition compared to other jet observables and thus allows for accurate theoretical predictions from higher-order computations. Secondly, the energy correlator can be defined both perturbatively and non-perturbatively, which makes it an ideal observable to study non-perturbative power corrections. Particularly, it connects the QFT operators and CFT techniques to phenomenological studies and real-world experiments (see Refs.~\cite{Chen:2020adz,Chen:2021gdk,Chen:2022jhb,Chen:2023wah,Chen:2023zzh,Chen:2024nyc}). Lastly, there are various infrared limits in the multi-point energy correlators as discussed in Ref.~\cite{Yang:2022tgm}, such as collinear, squeezed, and coplanar limits. It is a playground for developing EFT techniques and applying them to precision QCD measurements.

In this paper, we propose a new projection $\tau_p$ of the three-point energy correlator and study the coplanar limit from the $\tau_p\to 0$ region. In particular, we focus on the so-called trijet configuration where there are three well-separated jets falling on the same plane, and to guarantee infrared safety, we identify these three jets by the $k_T$ algorithm. From trijet kinematics of $e^+e^-\to\gamma^*\to q \bar q g$, i.e. the effects of collinear splitting and soft recoil, we derive a TMD-based factorization in SCET to describe the large logarithms enhancement $\ln(\tau_p)$ in this region. It turns out to be a natural generalization of the back-to-back factorization for a two-point correlator, EEC, where the collinear dynamics are captured by the TMD fragmentation function, and the soft emissions are organized into the trijet TMD soft function. 

The only missing ingredient to achieve N${}^3$LL resummation is the two-loop $q\bar q g$ hard function, which we extract from the QCD form factors in the literature. Then we solve the RGEs for individual ingredients and perform the resummation for coplanar EEEC in the $\gamma^* \to q\bar q g$ channel. To our best knowledge, this is the first trijet observable that reaches the N${}^3$LL accuracy. We also estimate the perturbative uncertainties by varying all the renormalization scales systematically and find a good convergence from NLL to N${}^3$LL. This is the main result of the paper. 
For completeness, we also look into the resummation for the $\gamma^*\to ggg$ channel. 
Since it is $\cO(\alpha_s^2)$ suppressed relative to the $\gamma^*\to q\bar qg$ channel, 
one should add the N${}^{k-2}$LL resummation as correction to the $q\bar q g$ channel at N${}^k$LL accuracy.
We therefore compute it to NLL and compare it with N$^3$LL $q\bar q g$ channel.
We also emphasize that this is the first resummation program for the ``light-by-glue'', $\gamma^*\to ggg$ channel.

In Refs.~\cite{Yan:2022cye,Yang:2022tgm,Yang:2024gcn}, we proposed another way of probing the coplanar limit, i.e. $s\to 1$ limit under the celestial parameterization, which is referred to as fully differential spectrum in this paper. We generalize our $\tau_p$ factorization to the fully differential case and also perform the N${}^3$LL resummation for $\ln(1-s)$ logarithms for fixed trijet shapes (i.e. fixed $\phi_1$ and $\phi_2$). For a generic value of $\phi_{1,2}$, we find similar distributions as the $\tau_p$ projection and good convergence. However, when approaching the squeezed limit, we receive additional enhancement from logarithms of $\phi_{1,2}$ and they also need to be resummed. It will be interesting to study the factorization for the squeezed limit and explore the joint resummation with the coplanar limit. 

As mentioned in Sec.~\ref{sec:factorization}, the factorization theorem and resummation we derived can be applied to other processes, such as Higgs decays and Z decays, and the only required change is the hard function. Although the data from the Large Electron-Position Collider might not have enough trijet events to perform these analyses, future colliders like CEPC, ILC and FCC-ee will allow the precise measurements of coplanar EEEC. On the one hand, the more inclusive spectrum, $\tau_p$ projection, can be applied to the extractions of Standard Model parameters, e.g. the strong coupling constant $\alpha_s$, Higgs mass and Yukawa couplings. On the other hand, the fully differential spectrum provides access to different hard kinematics and can contribute to removing the QCD backgrounds in the data. Moreover, it will be interesting to apply our factorization to the boosted top decay, where the decay particles will be formed into three QCD jets on the same plane in the top rest frame. The coplanar resummation will then improve the theoretical prediction of the Jacobian peak of the top decay EEEC and help with the top quark mass extraction at the LHC. 
\begin{acknowledgments}
We would like to thank Hua Xing Zhu for collaboration at the early stage of this project.
We also thank Arindam Bhattacharya, Miguel Benitez, Kyle Lee, Johannes Michel, Ian Moult, Matthew Schwartz, Iain Stewart and Hua Xing Zhu for useful discussions.
A.G. was supported by the U.S.~Department of Energy, Office of Science, Office of Nuclear Physics, from DE-SC0011090. T.-Z.Y wants to acknowledge the European Research Council (ERC) for the funding under the European Union's Horizon 2020 research and innovation programme grant agreement 101019620 (ERC Advanced Grant TOPUP). XY.Z. was supported in part by the U.S. Department of Energy under contract DE-SC0013607.
\end{acknowledgments}

\appendix

\section{Running coupling and RG kernels}
\label{app:betaRG}

In this section, we summarize the running coupling and the RG kernels in the resummation setup. The beta function is given as follows
\begin{equation}
	\label{eq:beta}
	\frac{\mathrm{d}\alpha_s(\mu)}{\mathrm{d}\ln \mu}= \beta (\alpha_s(\mu)),\quad 
	\beta (\alpha)=- 2 \alpha\, \left[ \left( \frac{\alpha}{4 \pi} \right) \beta_0 + \left(
	\frac{\alpha}{4 \pi} \right)^2 \beta_1 + \left( \frac{\alpha}{4 \pi}
	\right)^3 \beta_2 + \cdots \right]\,,
\end{equation}
where the coefficient up to three loops are given by~\cite{Tarasov:1980au,Larin:1993tp,vanRitbergen:1997va,Czakon:2004bu}
\begin{align}
	\beta_0 &=\frac{11}{3} C_A - \frac{4}{3} T_F n_f 
	\,, \\
	\beta_1 &= \frac{34}{3} C_A^2 - \frac{20}{3} C_A T_F n_f - 4 C_F T_F n_f 
	\nn \,,\\
	\beta_2 &= n_f^2 T_F^2 \left(\frac{158 }{27}C_A+\frac{44
	}{9}C_F\right)+n_f T_F \left(2
	C_F^2-\frac{205 }{9}C_FC_A-\frac{1415 }{27}C_A^2\right)+\frac{2857 }{54}C_A^3
	\,,\nn\\
	\beta_3 &= \frac{1093}{729} n_f^3
	+\left(\frac{50065}{162} + \frac{6472}{81}\zeta_3\right) n_f^2
	+\left(-\frac{1078361}{162} - \frac{6508}{27} \zeta_3 \right) n_f + 3564 \zeta_3 + \frac{149753}{6} \nn\,.
\end{align}

At one-loop, Eq.~\eqref{eq:beta} can be solved exactly. At two-loop and beyond, there are different solutions. In this paper, we use the so-called iterative solution, where the RGE is solved iteratively order-by-order in the expansion of $\beta_n/\beta_0$. Explicitly, 
the four-loop running coupling needed for the N${}^3$LL resummation is
\begin{align}
    \alpha_s(\mu) &= \alpha_s(Q)\Bigg\{X+\frac{\alpha_s(Q)}{4\pi }\frac{\beta_1}{\beta_0}\ln X+\frac{\alpha_s^2(Q)}{16\pi^2}\left[\frac{\beta_2}{\beta_0}\left(1-\frac{1}{X}\right)+\frac{\beta_1^2}{\beta_0^2}\left(\frac{1}{X}-1+\frac{\ln X}{X}\right)\right]\nn\\
    &+\frac{\alpha_s^3(Q)}{64\pi^3}\left[\frac{\beta_3}{2\beta_0}\left(1-\frac{1}{X^2}\right)+\frac{\beta_1\beta_2}{\beta_0^2}\left(\frac{1-X}{X}+\frac{\ln X}{X^2}\right)+\frac{\beta_1^3}{2\beta_0^3}\left(1+\frac{1}{X^2}-\frac{2}{X}-\frac{\ln^2 X}{X^2}\right)\right]\Bigg\}^{-1}
\end{align}
where
\begin{equation}
    X\equiv 1+\frac{\alpha_s(Q)}{2\pi} \beta_0\ln\frac{\mu}{Q}
\end{equation}
Note that the two-loop and three-loop running needed for lower-order resummation can be obtained by only keeping the $\mathcal{O}(\alpha_s)$ and $\mathcal{O}(\alpha_s^2)$ terms inside the curly bracket. For the numerical calculation, we choose $\alpha_s(m_Z)=0.118$. 

In the main text, we define the RG kernels in Eq.~\eqref{eq:RGkernel_def} and use them to express the resummed formula. Here we also provide the explicit expressions at different orders~\cite{Bosch:2003fc,Neubert:2004dd}. For $S_\Gamma(\nu,\mu)$, we have
\begin{equation}
	S_\Gamma(\nu,\mu)=S_\Gamma^{\text{LL}}(\nu,\mu)+S_\Gamma^{\text{NLL}}(\nu,\mu)+S_\Gamma^{\text{NNLL}}(\nu,\mu)+\cdots
\end{equation}
with 
\begin{align}
	S_\Gamma^{\text{LL}}(\nu,\mu)&=\frac{\Gamma_0\pi}{\beta_0^2 \alpha_s(\nu)}\left(1-\frac{1}{r}-\ln r\right)
    \,, \notag\\
	S_\Gamma^{\text{NLL}}(\nu,\mu)&=\frac{\Gamma_0}{4\beta_0^2}\left[\left(\frac{\Gamma _1}{\Gamma _0}-\frac{\beta
   _1}{\beta _0}\right) (1-r+\ln r)+\frac{\beta
   _1 }{2 \beta _0}\ln^2 r\right]
   \,, \notag \\
   S_\Gamma^{\text{NNLL}}(\nu,\mu)&=\frac{\Gamma_0 \alpha_s(\nu)}{32\pi\beta_0^2}\bigg[\left(\frac{\beta _1^2}{\beta _0^2}-\frac{\beta
   _2}{\beta _0}\right) \left(1-r^2+2 \ln r\right)+ \left(\frac{\beta _1 \Gamma
   _1}{\beta _0 \Gamma _0}-\frac{\Gamma _2}{\Gamma
   _0}\right)(1-r)^2
   \notag \\
   &+2 \left(\frac{\beta _1 \Gamma
   _1}{\beta _0 \Gamma _0}-\frac{\beta _1^2}{\beta
   _0^2}\right) (1-r+r \ln r)\bigg]\nn\\
   S_\Gamma^{\text{N}^3\text{LL}}(\nu,\mu)&=\frac{\Gamma_0 \alpha_s(\nu)^2}{64\pi^2\beta_0^2}\Bigg[\left(\frac{\beta
   _2}{\beta _0}-\frac{\beta _1^2}{\beta _0^2}\right)\left(\frac{\Gamma _1}{\Gamma _0}-\frac{\beta
   _1}{\beta _0}\right)\frac{(1-r)^2(2+r)}{3}+\left(\frac{\Gamma_3}{\Gamma_0}-\frac{\beta_3}{\beta_0}-\frac{\beta_1\Gamma_2}{\beta_0\Gamma_0}+\frac{\beta_1\beta_2}{\beta_0^2}\right)\nn\\
   &\left(\frac{1-r^3}{3}-\frac{1-r^2}{2}\right)+\frac{\beta_1}{\beta_0}\left(\frac{\Gamma_2}{\Gamma_0}-\frac{\beta_2}{\beta_0}-\frac{\beta_1\Gamma_1}{\beta_0\Gamma_0}+\frac{\beta_1^2}{\beta_0^2}\right)\left(\frac{1-r^2}{4}+\frac{r^2\ln r}{2}\right)\nn\\
   &+\left(-\frac{\beta_3}{\beta_0}+\frac{2\beta_1\beta_2}{\beta_0^2}-\frac{\beta_1^3}{\beta_0^3}\right)\left(\frac{1-r^2}{4}+\frac{\ln r}{2}\right)\Bigg]\,.
   \end{align}
Here we define $r\equiv \frac{\alpha_s(\mu)}{\alpha_s(\nu)}$. Similarly, we have
\begin{align}
	A_\Gamma(\nu,\mu)=A_\Gamma^{\text{LL}}(\nu,\mu)+A_\Gamma^{\text{NLL}}(\nu,\mu)+A_\Gamma^{\text{NNLL}}(\nu,\mu)+\cdots
	\,, \notag \\
	A_{\gamma_j}(\nu,\mu)=A_{\gamma_j}^{\text{LL}}(\nu,\mu)+A_{\gamma_j}^{\text{NLL}}(\nu,\mu)+A_{\gamma_j}^{\text{NNLL}}(\nu,\mu)+\cdots
\,.\end{align}
The corresponding term at each order is
\begin{align}
	A_\Gamma^{\text{LL}}(\nu,\mu)&=\frac{\Gamma_0}{2\beta_0}\ln r
	\,, \notag \\
	A_\Gamma^{\text{NLL}}(\nu,\mu)&=\frac{\Gamma_0\alpha_s(\nu)}{8\pi\beta_0}\left(\frac{\beta_1}{\beta_0}-\frac{\Gamma_1}{\Gamma_0}\right)(1-r)
	\,, \notag \\
	A_\Gamma^{\text{NNLL}}(\nu,\mu)&=\frac{\Gamma _0  \alpha (\nu )^2}{64 \pi ^2 \beta _0}\left(\frac{\beta _1^2}{\beta _0^2}-\frac{\beta
   _2}{\beta _0}+\frac{\Gamma _2}{\Gamma_0}-\frac{\beta _1 \Gamma _1}{\beta _0 \Gamma_0}\right)\left(r^2-1\right)\,,\nn\\
   A_\Gamma^{\text{N}^3\text{LL}}(\nu,\mu)&=\frac{\Gamma _0  \alpha (\nu )^3}{384 \pi ^3 \beta _0}\left[\frac{\Gamma_3}{\Gamma_0}-\frac{\Gamma_2\beta_1}{\Gamma_0\beta_0}+\frac{\Gamma_1}{\Gamma_0}\left(\frac{\beta_1^2}{\beta_0^2}-\frac{\beta_2}{\beta_0}\right)-\frac{\beta_1^3}{\beta_0^3}+\frac{2\beta_1\beta_2}{\beta_0^2}-\frac{\beta_3}{\beta_0}\right](1-r^3)
\end{align}

Expressions for $A_{\gamma_j}(\nu,\mu)$ are similar with the replacement of anomalous dimensions.

\section{Anomalous dimensions}
\label{app:anomalous_dimensions}

In this section, we summarize all the anomalous dimensions needed for the resummation. First of all, the cusp anomalous dimension has been computed up to four loops in Ref.~\cite{Korchemsky:1987wg, Moch:2004pa,Moch:2017uml, Moch:2018wjh, Davies:2016jie,Henn:2019swt,vonManteuffel:2020vjv}.

\begin{align} \gamma_{\text{cusp}}(\alpha_s)&=\left(\frac{\alpha_s}{4\pi}\right)\Gamma_0+\left(\frac{\alpha_s}{4\pi}\right)^2 \Gamma_1+\left(\frac{\alpha_s}{4\pi}\right)^3 \Gamma_2+\cdots\notag 
 \,,
\end{align}
where \cite{Korchemsky:1987wg,Moch:2004pa}
\begin{align}
    \Gamma_0&=4 
      \,, \notag\\
	\Gamma_1&=4\left[C_A\left(\frac{67}{9}-\frac{\pi^2}{3}\right)-\frac{20}{9}T_F n_F\right]
      \,, \notag\\
	\Gamma_2&=4 \bigg[\left(-\frac{56 \zeta_3}{3}-\frac{418}{27}+\frac{40 \pi ^2}{27}\right)
   C_A n_f T_F+\left(\frac{22 \zeta_3}{3}+\frac{245}{6}-\frac{134 \pi ^2}{27}+\frac{11
   \pi ^4}{45}\right) C_A^2
      \notag\\
   &\qquad+\left(16 \zeta_3-\frac{55}{3}\right) C_F n_f T_F-\frac{16}{27}
   n_f^2 T_F^2\bigg]\,,\nn\\
   \Gamma_{3,q} &=    15526.5-3878.93 n_f +146.683  n_f^2+ 2.454 n_f^3\,,\nn\\
   \Gamma_{3,g}  &= 13626.7-3904.67 n_f+146.683  n_f^2+ 2.454  n_f^3\,.
\end{align}
For four-loop cusp $\Gamma_3$, we only provide the numerical values and the full analytic result can be found in \cite{Henn:2019swt,vonManteuffel:2020vjv}. Note that the Casimir scaling between quark and gluon cusp anomalous dimensions is broken at four-loop, and thus we give the values for both of them.

For the regular anomalous dimensions $\gamma[\alpha_s, \ldots]$, with the dots representing potential dependence on kinematical variables, we also expand them in $\alpha_s/(4\pi)$,
\begin{align}
\gamma[\alpha_s,\ldots] = \sum_{n=0}^\infty \left(\frac{\alpha_s}{4 \pi} \right)^{n+1} \gamma_n[\ldots]   \,.
\end{align}
The quark and gluon anomalous dimensions through to three loops are \cite{Moch:2005id,Moch:2005tm,Idilbi:2005ni,Idilbi:2006dg,Becher:2006mr}
\begin{align}
  \gamma^{q}_0 = & \, -3 C_F \,,
\nbrk 
  \gamma^{q}_1 = & \, C_A C_F \left(-11 \zeta_2+26
    \zeta_3-\frac{961}{54}\right)
+C_F^2 \left(12 \zeta_2-24 \zeta_3-\frac{3}{2}\right)+C_F n_f \left(2 \zeta_2+\frac{65}{27}\right) \,,
\nbrk
  \gamma_2^{q} = & \,   
     -\frac{4880 \pi ^2 \zeta _3}{81}+\frac{82072 \zeta _3}{27}-\frac{15328 \zeta
   _5}{9}-\frac{2066 \pi ^4}{405}-\frac{5062 \pi
   ^2}{81}-\frac{196621}{243}
   \nbrk &
   +\left(-\frac{7472 \zeta _3}{81}+\frac{68 \pi ^4}{1215}+\frac{4564 \pi
   ^2}{243}+\frac{36236}{729}\right) n_f+\left(-\frac{32 \zeta _3}{81}-\frac{40 \pi ^2}{81}+\frac{9668}{2187}\right)
   n_f^2 \,,
\nbrk
  \gamma_0^{g} = &\, - \beta_0 \,,
\nbrk
  \gamma_1^{g} = &\, \,C_A^2 \left(\frac{11 \zeta_2}{3}+2 \zeta_3-\frac{692}{27}\right)+C_A n_f \left(\frac{128}{27}-\frac{2 \zeta_2}{3}\right)+2 C_F n_f \,,
  \nbrk
  \gamma_2^{g} = &\,  -60 \pi ^2 \zeta _3+1098 \zeta _3-432 \zeta
   _5-\frac{319 \pi ^4}{10}+\frac{6109 \pi ^2}{18}-\frac{97186}{27}
   \nbrk &
  +\left(\frac{460 \zeta _3}{9}+\frac{107 \pi ^4}{45}-\frac{635 \pi
   ^2}{27}+\frac{59635}{162}\right) n_f+\left(-\frac{56 \zeta _3}{9}+\frac{10 \pi ^2}{27}-\frac{1061}{486}\right)
   n_f^2 \,.
\end{align}
Next, The soft anomalous dimension up to three loops is~\cite{Li:2014afw}
\begin{align}
    \gsoft_{0} &=  \, 0\,,
\nbrk
  \gsoft_{1} &= \, C_A \left(-\frac{808}{27}+\frac{22}{3}\zeta_2+28 \zeta_3\right)+ n_f \left(\frac{112}{27}-\frac{4}{3}\zeta_2\right)\,,
\nbrk
  \gsoft_{2} &=  \, C_A^2 \left(-\frac{136781}{729}+\frac{12650}{81}\zeta_2+\frac{1316}{3}\zeta_3-176\zeta_4-192 \zeta_5-\frac{176}{3} \zeta_3\zeta_2\right)
\brk
+C_A n_f \left(\frac{11842}{729}-\frac{2828}{81}\zeta_2-\frac{728}{27}\zeta_3+48 \zeta_4\right) \brk
+C_F n_f \left(\frac{1711}{27}-4\zeta_2-\frac{304}{9}\zeta_3-16 \zeta_4\right)
+ n_f^2 \left(\frac{2080}{729}+\frac{40}{27}\zeta_2-\frac{112}{27}\zeta_3\right)\,.
\label{eq:c2}
\end{align}
Finally, the rapidity anomalous dimension up to three loops is \cite{Li:2016ctv} 
\begin{align}
  \label{eq:8}
  \ugr_0 &=  \, 0\,,
\nbrk
\ugr_1 &= \, C_A  \left(-\frac{808}{27}+28 \zeta_3\right)+ n_f\frac{112}{27}\,,
\nbrk
\ugr_2 &= \, C_A^2 \left(-\frac{297029}{729} +\frac{6392}{81}\zeta_2+\frac{12328}{27} \zeta_3+\frac{154}{3}\zeta_4-192\zeta_5-\frac{176}{3} \zeta_3\zeta_2 \right) \brk
+ C_A n_f \left(\frac{62626}{729}-\frac{824}{81}\zeta_2-\frac{904}{27}\zeta_3+\frac{20}{3}\zeta_4
\right) 
+ n_f^2 \left(-\frac{1856}{729}-\frac{32}{9} \zeta_3 \right)
\brk
+ C_F n_f \left(\frac{1711}{27} -\frac{304}{9}\zeta_3 -16 \zeta_4 \right)\,.
\end{align} 

\section{Fourier transformation}\label{app:fouriertransf}

In this section, we derive the analytic transformation maps between logarithms in the Fourier and momentum space. Given the form in Eq.~\eqref{eq:Finalresum}, we consider the following Fourier transformation convention
\begin{equation}
    \mathcal{F}\left[\left(\frac{b_y \mu}{b_0}\right)^{2\epsilon}\right]\equiv\int_{-\infty}^{+\infty}\frac{db_y}{2\pi\xi}\,2 \cos\left(\frac{b_y \tau_p}{\xi}\right)\left(\frac{b_y \mu}{b_0}\right)^{2\epsilon}=-\frac{2\Gamma(1+2\epsilon)\sin(\pi\epsilon)}{\pi\tau_p}\left(\frac{\mu\xi}{b_0\tau_p}\right)^{2\epsilon}
\end{equation}
For convenience, we will define $x=\frac{2\tau_p}{\mu\xi}$, $L=\ln\left(\frac{b_y^2\mu^2}{b_0^2}\right)$ and expand both sides in $\epsilon$. Note that the right-hand side gives rise to the plus distributions in x while the left-hand side contains powers of $L$. By mapping the coefficient in the same order of $\epsilon$, we obtain the analytic expressions for the Fourier transformation relevant to the fixed-order expansion. The first few orders read
\begin{align}
    \mathcal{F}[1]&=\frac{2}{\mu\xi}\,\delta\left(x\right)\,,\nn\\
    \mathcal{F}\left[L\right]&=-\frac{4}{\mu\xi}\left[\frac{1}{x}\right]_{+}\,,\nn\\
    \mathcal{F}\left[L^2\right]&=\frac{2\pi^2}{3\mu\xi}\delta(x)+\frac{16}{\mu\xi}\left[\frac{\ln(x)}{x}\right]_{+}\,,\nn\\
    \mathcal{F}\left[L^3\right]&=-\frac{48}{\mu\xi}\left[\frac{\ln^2(x)}{x}\right]_{+}-\frac{4\pi^2}{\mu\xi}\left[\frac{1}{x}\right]_{+}-\frac{32\zeta_3}{\mu\xi}\delta(x)\,,\nn\\
    \mathcal{F}\left[L^4\right]&=\frac{128}{\mu\xi}\left[\frac{\ln^3(x)}{x}\right]_{+}+\frac{32\pi^2}{\mu\xi}\left[\frac{\ln(x)}{x}\right]_{+}+\frac{256\zeta_3}{\mu\xi}\left[\frac{1}{x}\right]_{+}+\frac{38\pi^4}{15\mu\xi}\delta(x)\,,\nn\\
    \mathcal{F}\left[L^5\right]&=-\frac{320}{\mu\xi}\left[\frac{\ln^4(x)}{x}\right]_{+}-\frac{160\pi^2}{\mu\xi}\left[\frac{\ln^2(x)}{x}\right]_{+}-\frac{2560\zeta_3}{\mu\xi}\left[\frac{\ln(x)}{x}\right]_{+}-\frac{76\pi^4}{3\mu\xi}\left[\frac{1}{x}\right]_{+}\nn\\
    &-\frac{64(5\pi^2\zeta_3+72\zeta_5)}{3\mu\xi}\delta(x)
    \,.
\end{align}


\section{Fixed-order expansion}\label{app:resum_fo}

Here we provide the fixed-order expansion of the resummed formula in Eq.~\eqref{eq:Finalresum}.
\begin{align}
    \frac{1}{\sigma_{\text{LO}}}\frac{d\sigma_q}{d\tau_p}&=\int_T \df v \df w H_q^{(0)}(v,w) \left[\left(\frac{\alpha_s}{4\pi}\right)A_1(v,w,\tau_p)+\left(\frac{\alpha_s}{4\pi}\right)^2 A_2(v,w,\tau_p)+\mathcal{O}(\alpha_s^3)\right]
\end{align}
where the LO singular coefficient is
\begin{align}
    A_1(v,w,\tau_p)&=-8 \left(C_A+2 C_F\right)\frac{\ln \left(\frac{2 \tau _p}{\xi  Q}\right)}{\tau_p}\nn\\
    &+\left(4 C_A \ln \left(\frac{v w}{1-v-w}\right)-\frac{22 C_A}{3}+8 C_F \ln (1-v-w)-12 C_F+\frac{8 T_F
   n_F}{3}\right)\frac{1}{\tau_p}
\end{align}
and the NLO singular coefficient is
\begin{align}
    A_2(v,w,\tau_p)&=32(C_A+2C_F)^2\frac{\ln^3 \left(\frac{2 \tau _p}{\xi  Q}\right)}{\tau_p}-(C_A+2C_F)\left[48 C_A \ln \left(\frac{v w}{1-v-w}\right) \right.\nn\\
    &\left. +96 C_F \ln (1-v-w)-\frac{440}{3}C_A-144C_F+\frac{160}{3}n_f T_F\right]\frac{\ln^2 \left(\frac{2 \tau _p}{\xi  Q}\right)}{\tau_p}\nn\\
    &+\Bigg[-8(C_A+2C_F)\frac{H_1(v,w)}{H_0(v,w)}+16C_A^2 \ln^2\left(\frac{vw}{1-v-w}\right)\\
    & +\left(64 C_F C_A \ln (1-v-w)-88 C_A^2-96C_F C_A +32 C_A T_F n_f \right)\ln\left(\frac{vw}{1-v-w}\right)\nn\\
    &+64 C_F^2 \ln^2 (1-v-w)-\left(176 C_F C_A+192 C_F^2-64C_F T_F n_f\right)\ln(1-v-w)\nn\\
    &+\left(\frac{68}{3}C_A^2+\frac{256}{3}C_F C_A +80C_F^2\right)\pi^2+\frac{172}{9}C_A^2+\frac{208}{9}C_F C_A+16 C_F^2-56 C_A T_F n_f\nn\\
    &-\frac{464}{9} C_F T_F n_f+\frac{128}{9} n_f^2 T_F^2\Bigg]\frac{\ln \left(\frac{2 \tau _p}{\xi  Q}\right)}{\tau_p}\nn\\
    &+\Bigg[\frac{H_1(v,w)}{H_0(v,w)} \left(4C_A\ln\left(\frac{vw}{1-v-w}\right) +8C_F \ln(1-v-w)-\frac{22}{3}C_A-12C_F+\frac{8}{3}T_F n_f\right)\nn\\
    &+\left(\left(-\frac{34}{3}C_A^2-20C_F C_A\right)\pi^2+\frac{389}{9}C_A^2+32 C_F C_A-\frac{100}{9}C_A T_F n_f\right)\ln\left(\frac{vw}{1-v-w}\right)\nn\\
    &+\left(  \left(-40 C_F^2-\frac{68}{3}C_F C_A\right)\pi^2+64 C_F^2 +\frac{796}{9} C_F C_A-\frac{200}{9}C_F T_F n_f  \right)\ln(1-v-w)\nn\\
    &+\left(40 C_A^2+304 C_F C_A+160 C_F^2\right)\zeta_3+\left(\frac{319}{9}C_A^2+\frac{820}{9}C_F C_A+68C_F^2-\frac{116}{9}C_A T_F n_f\right.\nn\\
    &\left. -\frac{200}{9}C_F T_F n_f\right)\pi^2-\frac{2006}{27}C_A^2-172C_F C_A-102 C_F^2+\frac{1028}{27}C_A T_F n_f+60 C_F T_F n_f\nn \\
    & -\frac{80}{27}n_f^2 T_F^2\Bigg]\frac{1}{\tau_p}
    \,.
\end{align}

Note that the $H_q^{(0)}(v,w)$ and $H_q^{(1)}(v,w)$ are the tree-level and one-loop hard function. $H_q^{(0)}(v,w)$ is given in Eq.~\eqref{eq:hard0qqg}, and $H_q^{(1)}(v,w)$ is provided in the ancillary file.

\bibliography{Coplanar}{}
\bibliographystyle{JHEP}
\end{document}